\documentclass[11pt]{article}

\usepackage[T1]{fontenc}
\usepackage[utf8]{inputenc}
\usepackage{amsmath}
\usepackage{newtxtext,newtxmath}
\usepackage{microtype}

\usepackage[
    letterpaper,
    left=1.0in,
    right=1.0in,
    top=0.95in,
    bottom=1.0in,
    headheight=14pt,
    headsep=14pt,
    footskip=24pt
]{geometry}

\usepackage{setspace}
\usepackage{fancyhdr}
\usepackage{authblk}

\usepackage[version=4]{mhchem}
\usepackage{siunitx}

\usepackage{graphicx}
\graphicspath{{PNG}{Figs/}}
\usepackage{subcaption}
\usepackage{booktabs}
\usepackage{multirow}
\usepackage{tikz}
\usepackage{caption}
\usepackage[english,linesnumbered,algoruled]{algorithm2e}
\usepackage[noend]{algpseudocode}

\usepackage{enumitem}
\setlist{nosep}

\usepackage{xcolor}
\definecolor{linkcol}{rgb}{0,0,0.4}
\definecolor{citecol}{RGB}{0,20,95}

\usepackage[
    colorlinks=true,
    linkcolor=linkcol,
    citecolor=citecol,
    urlcolor=linkcol,
    bookmarksopen=true,
    pdftitle={Probabilistic Upscaling of Hydrodynamics in Geological Fractures Under Uncertainty},
    pdfauthor={Sarah Perez et al.}
]{hyperref}
\usepackage[mathlines]{lineno}

\usepackage{titlesec}
\titleformat{\section}{\Large\bfseries}{\thesection}{0.6em}{}
\titleformat{\subsection}{\large\bfseries}{\thesubsection}{0.6em}{}
\titlespacing*{\section}{0pt}{1.2em}{0.5em}
\titlespacing*{\subsection}{0pt}{1.0em}{0.4em}

\newcommand{\OmegaD}[1]{\ensuremath{\Omega^{#1 D}}}

\title{\bfseries Probabilistic Upscaling of Hydrodynamics\\
in Geological Fractures Under Uncertainty}

\author[1,3]{Sarah Perez\thanks{Corresponding author: \href{mailto:S.Perez@hw.ac.uk}{S.Perez@hw.ac.uk}}}
\author[2,3]{Florian Doster}
\author[2,3]{Hannah Menke}
\author[2,3]{Ahmed ElSheikh}
\author[1,3]{Andreas Busch}

\affil[1]{The Lyell Centre, Heriot-Watt University, Edinburgh, UK}
\affil[2]{Institute of GeoEnergy Engineering, Heriot-Watt University, Edinburgh, UK}
\affil[3]{Subsurface Energy Transition and Innovation Centre, Heriot-Watt University, Edinburgh, UK}

\begin{document}

\maketitle

\begin{abstract}
Flow and transport in fractured geological media are strongly controlled by aperture heterogeneity and uncertainty in subsurface characterisation, yet most upscaling approaches rely on deterministic representations of fracture permeability. This study presents a scalable probabilistic workflow that bridges image-based fracture geometry and uncertainty-aware hydraulic predictions across scales. The approach integrates Bayesian correction of aperture-permeability model misspecification, a deep learning surrogate for predicting spatially distributed permeability statistics, and Darcy-scale flow upscaling to propagate uncertainty to effective transmissivity.

The workflow is applied to natural shear fractures from core material in the Little Grand Wash Fault damage zone (Utah) and to simplified geometries derived from the same $\mu$CT datasets. The Bayesian component quantifies epistemic uncertainty due to measurement errors and imperfect constitutive relations, while a Residual U-Net learns the effects of local heterogeneity and spatial correlation on predicted permeability uncertainty. Together, these components generate ensembles of permeability fields that are subsequently upscaled to probabilistic macroscopic flow responses.

Results show that common empirical aperture-permeability relations are systematically biased for natural fractures, whereas the proposed probabilistic workflow yields uncertainty-aware permeability estimates consistent with physics-based reference behaviour. The method captures the impact of channelisation, connectivity, and complex three-dimensional void geometries on transmissivity while quantifying the resulting uncertainty bounds. Computational efficiency arises from the proposed hybrid strategy for probabilistic upscaling, which combines physics-informed and data-driven approaches, preserves Stokes-flow consistency and supports uncertainty propagation without repeated high-fidelity simulations.\\

\end{abstract}
\vspace{-5mm}
\section*{Plain Language Summary}

Fractures play an important role in controlling how water and other fluids move underground. Predicting how easily a fracture transmits fluid is difficult because fractures are highly irregular, and small variations in their openings can strongly affect fluid flow properties. However, these openings are difficult to measure accurately, and common methods often simplify fracture properties and do not reflect uncertainty in subsurface characterisation.

Here, we developed a framework that combines physical modelling and artificial intelligence to estimate how fluids flow through fractured rocks while accounting for uncertainty in measurements and models. Using detailed images of natural rock fractures, we first correct simplified flow models using a limited number of high accuracy simulations. We then use a deep learning model to capture how local fracture structures influence permeability patterns and their uncertainty, which are propagated to predict a range of flow capacities.

Our approach allows rapid evaluation of many possible fracture scenarios for uncertainty quantification without repeating expensive simulations. This supports improved assessment of risks and performance in applications such as geothermal energy production, geological carbon storage, or groundwater protection. 

\section{Introduction}

Fluid flow through fractured rock controls a wide range of subsurface processes, including groundwater circulation, geothermal energy extraction, or the integrity of geological carbon storage \cite{Maldaner_2019, CHEN2026101580, RIZZO2024}. In many of these applications, hydraulic behaviour results from the interplay between conductive fractures or damage zones, whose transmissivity emerges from multi-scale aperture variability, roughness, and connectivity of preferential flow pathways \cite{Viswanathan_2022, Philippe_Davy_PNAS_2024}. Accurately upscaling such heterogeneous fracture-scale hydraulics into effective properties remains a long-standing challenge in water resources and geosciences, particularly given the substantial uncertainties associated with subsurface fracture characterisation.

Fracture geometries are only partially observable and are commonly inferred from sparse and indirect data sources, including borehole imaging, production logs, and seismic analysis \cite{ALDHAFEERI200718, Seismic_frac, Hosseinzadeh2023}. As a result, key inputs for fracture and fracture network modelling, such as aperture fields, contact patterns and intersections, and hydraulic properties are affected by substantial epistemic uncertainty \cite{NEJADI201721, wang_sequential_2026}. This motivates the use of analogue information to constrain the plausible hydraulic structures: field analogues (\emph{e.g.}, outcrops) provide geometric statistics at the surface, while laboratory analogues derived from core material enable controlled access to fracture void space at the millimetre-to-centimetre scales \cite{RIZZO2024, RAMANDI20164}. Here, we adopt a laboratory-analogue perspective and focus on natural shear fractures from core material extracted in the damage zone of the Little Grand Wash Fault, Utah \cite{KAMPMAN_2014_22, KAMPMAN_2014_51, Phillips2024, AchuoDze2026}.

Quantifying how uncertainty in fracture geometry and hydraulic properties propagates across scales is central to risk assessment and performance prediction in fractured subsurface systems \cite{Pettersson_2025, Hannah_Lu_2025, Michael_Liem_2025}. Yet, even with detailed fracture-scale geometry, translating uncertain geometric observations into hydraulic property distributions remains challenging. A physically consistent approach involves resolving flow in the fracture void space to recover the effective properties, but the computational cost of such direct simulations is prohibitive for the repeated evaluations required for uncertainty propagation \cite{Landry_2012, Adenike_2015}. Consequently, practical modelling often relies on simplified constitutive relationships that map mechanical aperture to permeability, most commonly through power-law formulations \cite{Zimmerman1996}. Such mappings rely on restrictive assumptions (\emph{e.g.}, smooth parallel plates), whereas natural fractures exhibit roughness, contact zones, tortuosity, and channelisation. These features produce strong deviations from deterministic aperture-permeability relationships and can bias local permeability estimates and their upscaled counterparts \cite{CRANDALL2010, Auradou_2009, HE2021}. Moreover, ambiguity in the definition of aperture itself can introduce additional uncertainty, particularly in complex void geometries where multiple disconnected or overlapping openings coexist along the fracture-normal direction (\emph{e.g.}, trapped fragments, branching openings, or partially cemented regions) \cite{Konzuk_2004, Landry_2024}. Indeed, when projecting such complex 3D geometries onto an aperture representation, several aperture values may coexist at a single in-plane location, which we refer to as a multivalued aperture structure. As a result, the definition of a representative mechanical aperture becomes inherently non-unique. Different aperture definitions can therefore lead to distinct spatial patterns, which in turn lead to uncertainty in hydraulic properties. These challenges motivate uncertainty-aware formulations in which mechanical aperture is treated as an uncertain geometric descriptor rather than a unique hydraulic proxy.

Taken together, these challenges highlight the need for practical upscaling strategies that are physically consistent, computationally tractable, and able to propagate uncertainty in fracture geometry and aperture-permeability relationships into probabilistic hydraulic predictions. In previous work \cite{GRL_Perez_2025}, we framed the discrepancy between Stokes flow and empirical Cubic-Law models as a model-misspecification problem and introduced a physics-based Bayesian correction to infer posterior distributions of local permeability conditioned on mechanical aperture measurements. This approach yielded, on synthetic fractures, spatially resolved estimates of expected permeability and local uncertainty that are consistent with Stokes-flow behaviour. Here, we build on this framework to generalise it to more complex fracture systems, including large natural shear fractures, and to develop a fully probabilistic upscaling approach. Specifically, uncertainty propagation is achieved by combining: (i) physics-based Bayesian inference, (ii) deep-learning surrogate trained to predict permeability distributions from mechanical aperture patches using Bayesian posterior statistics as supervision targets, and (iii) Darcy flow-based upscaling to compute effective permeability distributions and associated uncertainty bounds. This workflow replaces repeated Stokes-based simulations with fast surrogate-based uncertainty propagation while inheriting the physical constraints embedded in the Bayesian correction. The resulting permeability fields provide spatially resolved probabilistic descriptors (\emph{e.g.}, expected value, mode, and confidence bounds), which are then translated into hydraulic response distributions, enabling uncertainty-aware permeability mapping over complex fracture domains at low computational cost.

Furthermore, a key aspect of applying this workflow to realistic natural fractures is the presence of multivalued aperture structures in the segmented void space, arising from multiple disconnected or overlapping openings, which introduces intrinsic uncertainty in the definition of a representative mechanical aperture. Consistent with our uncertainty-aware formulation, in which mechanical aperture is treated as an uncertain descriptor rather than a hydraulic proxy, we adopt a probabilistic geometric representation of the fracture opening. To translate complex 3D geometries to 2D probabilistic permeability models, we introduce a thickness-weighted aperture reduction that yields a consistent mechanical-aperture descriptor while explicitly accounting for the contribution of coexisting branches. This reduction preserves information on the spatial organisation of conductive regions and provides a statistically interpretable input for the permeability inference. We validate the framework on natural shear fractures and on synthetic geometries derived from the same dataset, enabling comparisons between multivalued fracture geometries and simplified connected-band representations. While this study focuses on shear fractures in sedimentary rocks from the Little Grand Wash Fault damage zone, the proposed framework is general and applicable to fractures in a wide range of geological settings, including crystalline and carbonate rocks, where roughness, contact zones, and channelisation similarly control hydraulic behaviour.

The main contributions of this study are threefold. First, we present a hybrid physics-informed and data-driven framework for probabilistic upscaling of hydrodynamics in geological fractures, consistent with Stokes-flow behaviour and designed for uncertainty propagation. Second, we demonstrate how local uncertainty in the aperture-permeability relationship manifests as spatially heterogeneous uncertainty patterns and propagates to uncertainty bounds on effective fracture permeability. Third, we discuss how the proposed approach can be integrated into larger-scale workflows, including discrete fracture network (DFN) models, thereby supporting uncertainty-aware predictions at the network scale in fractured subsurface systems.

The paper is organised as follows. Section~\ref{sec:local_uncertainties} summarises the physics-based Bayesian correction used to infer local permeability statistics from mechanical aperture observations. Section~\ref{subsec:UNET} describes the deep learning architecture and training procedure, Section~\ref{subsec:prob_maps} introduces the construction of probabilistic permeability fields from the learned model and Section~\ref{subsec:Darcy_upscaling} presents Darcy flow-based upscaling of the resulting probabilistic permeability descriptors. Section~\ref{sec:Geological_frac} introduces the real fracture datasets and the aperture reduction strategy used to obtain a consistent mechanical-aperture descriptor. Section~\ref{sec:Results} reports validation of the surrogate, probabilistic permeability maps, and uncertainty propagation to effective permeability for natural fractures. Finally, Section~\ref{sec:discussion} discusses generalisation, computational efficiency, and extensions to fracture networks.

\section{Learning probabilistic hydraulic responses from mechanical aperture distributions}
\label{sec:local_uncertainties}

We investigate a probabilistic framework to infer reliable hydraulic responses of rough-walled geological fractures, assuming that flow is governed by the steady, incompressible Stokes equation. Rather than solving deterministic fluid dynamics for each fracture configuration, we learn the statistical relationship between mechanical aperture distributions and local hydraulic properties under uncertainty.

Uncertainty arises from two main sources: (i) measurement errors in estimating mechanical apertures, derived for instance from X-ray CT imaging of core samples \cite{Aperture_CT_Huo, RAMANDI20164, Kurotori2023}, and (ii) modelling uncertainty introduced by empirical aperture-permeability relationships based on Cubic Law assumptions \cite{Konzuk_2004, CRANDALL2010, HE2021}. By learning permeability distributions directly from uncertain aperture data, we aim to characterize both the expected hydraulic response and its associated uncertainty at the local scale.

\subsection{A physics-based Bayesian inference} 
\label{subsec:Bayesian_inference}

Understanding the hydraulic behaviour of rough-walled fractures requires accounting for the mismatch between Stokes flow and simplified empirical models typically used in practice. In previous work \cite{GRL_Perez_2025}, we formulated this discrepancy as a model misspecification problem, where the permeability predicted by Cubic-Law relationships is treated as a biased approximation of the true Stokes-consistent effective permeability, denoted $K_{NS}\in \mathbb{R}$. A Bayesian framework was introduced to infer the posterior distribution of the local permeability field $K(x,y)$ at each spatial location $(x,y)\in \OmegaD{2}$, conditioned on mechanical aperture measurements while remaining physically consistent. In this formulation, uncertainty arises from both measurement errors in the mechanical aperture field and model misspecification in the aperture–permeability relationship. Their relative contributions are automatically balanced during the inference through an adaptive weighting between data fidelity and model discrepancy, allowing the dominant source of uncertainty to emerge from the data.

In this Bayesian framework, the mechanical aperture field $a_m(x,y)$, defined as the distance between two opposite fracture walls measured perpendicular to a reference plane, serves as the primary data observable $\mathcal{D}$. Instead of using $a_m(x,y)$ as a direct surrogate for the fracture hydraulic transmissivity, the Bayesian inversion infers the hydraulic aperture $a_h(x,y)$ that is consistent with Stokes flow. This inferred hydraulic aperture directly governs the physically meaningful permeability field and allows us to correct the systematic bias introduced by Cubic-Law assumptions.

The inversion therefore yields a posterior distribution of the permeability field $K(x,y)$, fully characterised by its posterior predictive mean $\mathbb{E}[K(x,y)\,|\,\mathcal{D}]$, and posterior predictive variance  $\mathrm{Var}[K(x,y)\,|\,\mathcal{D}]$. These descriptors respectively represent the expected permeability and the associated local uncertainty at each spatial location  $(x,y)\in \OmegaD{2}$. Notably, they both emerge from the inferred hydraulic aperture $a_h(x,y)$, not from empirical aperture-permeability formulas. They thus encode essential information on how geometric uncertainties and model misspecification influence local permeability and flow behaviour.

This approach has been validated on synthetic fracture geometries with increasing roughness in Perez et al. (2025) \cite{GRL_Perez_2025}. We showed that the uncertainty ranges produced by the Bayesian correction consistently deviate from Cubic-Law predictions while capturing the upscaled Stokes permeability. This confirms that the corrected permeability field $K(x,y)$ is both probabilistic at the local scale and consistent at the upscaled level, while incorporating roughness-induced uncertainties. We refer to Perez et al. (2025) \cite{GRL_Perez_2025} for a complete formulation and detailed analysis of this physics-based Bayesian correction of model misspecification in estimating fracture hydraulic responses.

\subsection{A representative training dataset}
\label{subsec:training_dataset}

To propagate uncertainties across scales and extend the Bayesian correction to more complex fracture systems, including natural fractures and fracture networks, we combine the physics-based local inference described above with a data-driven deep learning model. Specifically, the Bayesian framework is first applied on local aperture patches to generate posterior permeability statistics, which are then used as training targets for a Convolutional Neural Network (CNN) that learns to reproduce these quantities directly from mechanical aperture fields. This hybrid strategy separates the expensive physics-based inference from the scalable learning stage: the Bayesian inference is performed at the patch scale to generate training data, while the trained CNN subsequently acts as a fast surrogate model, replacing repeated Stokes-based inversions on larger domains.

To construct a representative training dataset, we generate a collection of local mechanical aperture patches extracted from 140 distinct realistic fracture realisations with controlled permeabilities, roughness, and spatial correlation lengths (see Section~\ref{subsec:fracture_comparison} for a detailed analysis). For each realisation, the mechanical aperture field $a_m(x,y)$ is discretised into patches of size $128 \times 128$ pixels. Each patch is then processed through the Bayesian inversion framework to obtain the corresponding posterior predictive mean $\mathbb{E}[K(x,y)]$ and variance $\mathrm{Var}[K(x,y)]$.

The permeability of a natural fracture is then represented as a positive random variable, assuming that $K(x,y)$ follows a log-normal distribution at each location:
\begin{equation}
    K(x,y)\sim \mathrm{Log}\,\mathcal{N}\!\left(\mu(x,y),\sigma^2(x,y)\right), \qquad \forall (x,y)\in\OmegaD{2},
    \label{eq:log_normal_K}    
\end{equation}
where the log-normal parameters are obtained from the posterior predictive moments via
\begin{align}
    \sigma(x,y)
    &= \sqrt{
    \ln\!\left(
    1 +
    \left(\frac{\sqrt{\mathrm{Var}[K(x,y)]}}{\mathbb{E}[K(x,y)]}\right)^{2}
    \right)}\\[2mm]
    \text{and } \mu(x,y)
    &= \ln\!\big(\mathbb{E}[K(x,y)]\big) - \frac{1}{2}\sigma^2(x,y).
    \label{eq:mu_sigma_def}
\end{align}
The resulting fields $\mu(x,y)$ and $\sigma(x,y)$ then form the training targets of the neural network. The objective of the CNN is thus to approximate the mapping
\begin{equation}
   a_m(x,y)\;\longmapsto\; \big(\mu(x,y),\sigma(x,y)\big),
\label{eq:am_mu_sigma_mapping}
\end{equation}
thereby learning physically consistent local permeability statistics that implicitly reflect the uncertainty structure inferred from the Bayesian model, arising from both model misspecification and measurement uncertainty.

\begin{figure}[t!]
    \centering
    \begin{subfigure}{.53\textwidth}
        \caption{}
        \centering
        \includegraphics[width=\linewidth]{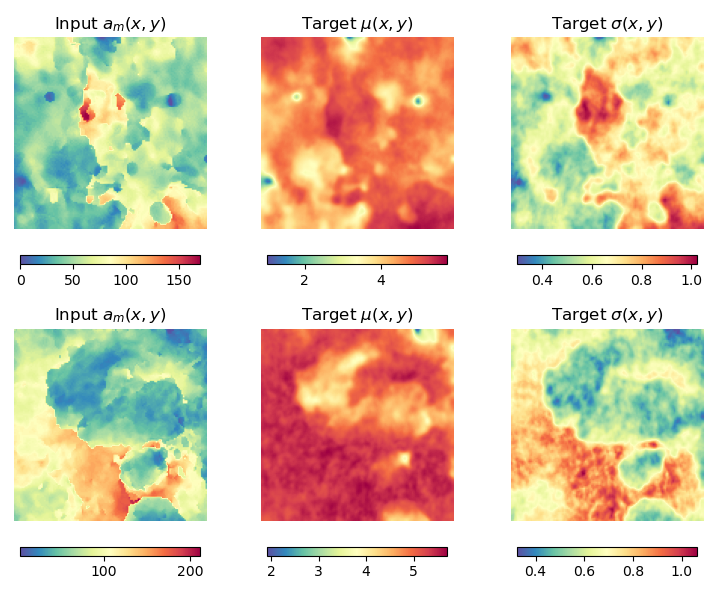}
    \end{subfigure}
    \hspace{-1mm}
    \begin{subfigure}{.45\textwidth}
        \caption{}
        \centering
        \includegraphics[width=\linewidth]{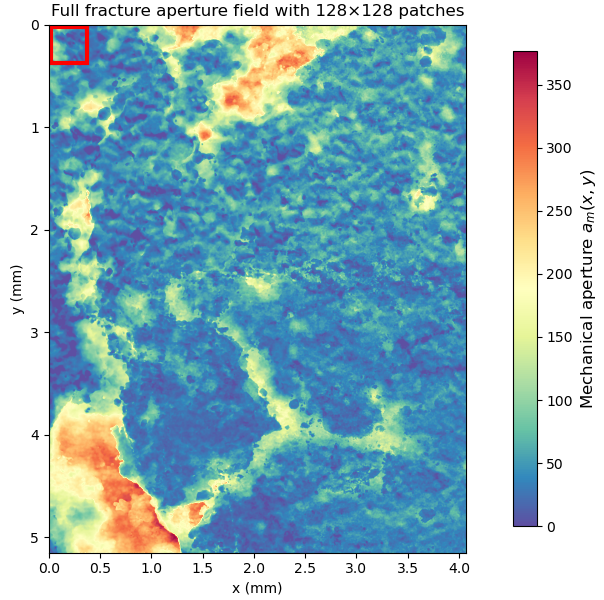}
    \end{subfigure}
    \caption{(a) Examples of paired input-target training patches used to learn the probabilistic permeability mapping \eqref{eq:am_mu_sigma_mapping}. The first column displays the mechanical aperture $a_m(x,y)$ in $\mu m$, while the second and third columns show the corresponding log-normal permeability parameters $\mu(x,y)$ and $\sigma(x,y)$. Together, these fields encode both the expected hydraulic response and its uncertainty at the patch scale, and serve as supervised learning targets for the CNN. (b) Example of a full mechanical aperture map of a target fracture, with a representative $128\times128$ patch highlighted by the red square to illustrate the spatial scale of the training samples relative to the full domain. This emphasises the local nature of the learning process and the scale separation between patch-based inference and fracture-scale predictions.
    }
\label{fig:Patch_samples}
\end{figure}

Data augmentation, including rotations and flips, is applied to increase variability and improve generalisation. The resulting 560 paired images $\{\,a_m(x,y), \, \big(\mu(x,y),\sigma(x,y)\big)\,\}$ capture a broad diversity of geological aperture geometries and permeability responses, enabling the network to learn a robust mapping from mechanical aperture distribution to log-normal permeability statistics. Because the Bayesian inference is performed under a prescribed flow direction (from right to left), the resulting permeability statistics $\mu(x,y)$ and $\sigma(x,y)$ implicitly inherit this orientation. To ensure that the CNN learns a physically consistent mapping, all mechanical aperture patches are oriented consistently with the prescribed flow direction prior to training. This guarantees that the learned mapping \eqref{eq:am_mu_sigma_mapping} reflects anisotropy and directional effects embedded in the training data, without imposing any constraints on how the surrogate model will later be used in the upscaling workflow. Figure~\ref{fig:Patch_samples} illustrates representative examples of the training patches, together with their corresponding target fields $\mu(x,y)$ and $\sigma(x,y)$. In Section~\ref{subsec:UNET}, we describe the U-Net architecture and training procedure used to efficiently learn this mapping.

\section{Uncertainty propagation across scale: a data-driven upscaling framework}
\label{sec:Up_fram}

We now introduce the data-driven upscaling framework designed to propagate local hydraulic uncertainties across scale by using a U-Net model based on Convolutional Neural Networks. By learning the probabilistic mapping \eqref{eq:am_mu_sigma_mapping} directly from data, the model captures how uncertain geometrical features and roughness patterns control the permeability distribution of natural fractures.

Because the training targets originate from the Bayesian inference described in Section~\ref{sec:local_uncertainties}, the network inherits the underlying Stokes physics while providing a scalable surrogate for uncertainty-aware predictions at larger scales. In the following, we detail the U-Net training strategy and the complete upscaling workflow used to assess the impact of uncertainty across scales.

\subsection{Patch-based learning with Residual U-Net}
\label{subsec:UNET}

Predicting permeability statistics on geological fracture systems requires a deep learning model capable of extracting spatial features across multiple scales. Natural fractures exhibit multi-scale heterogeneities including contact areas, channelisation and roughness-induced flow restrictions \cite{Meheust_2001, Kottwitz_2020, Landry_2024}. Both fine-scale asperities and larger connectivity patterns influence the hydraulic response depending on the spatial correlation lengths \cite{Meheust_2003, MAKEDONSKA_2016}. We therefore adopt a U-Net architecture \cite{Ronneberger2015UNet}, which combines a contracting encoder for hierarchical feature extraction and an expansive decoder reconstructing high-resolution fields through skip connections. This structure is widely used for pixel-wise prediction in porous and fractured media since it preserves fine-scale detail while retaining large-scale geometrical context \cite{TAKBIRI_2022, PHAM_2023, Yu_2023, WU_2024}.

\begin{figure}[t!]
\centering
\includegraphics[width=\linewidth,trim={0.1cm 1.2cm 0.1cm 1.5cm},clip]{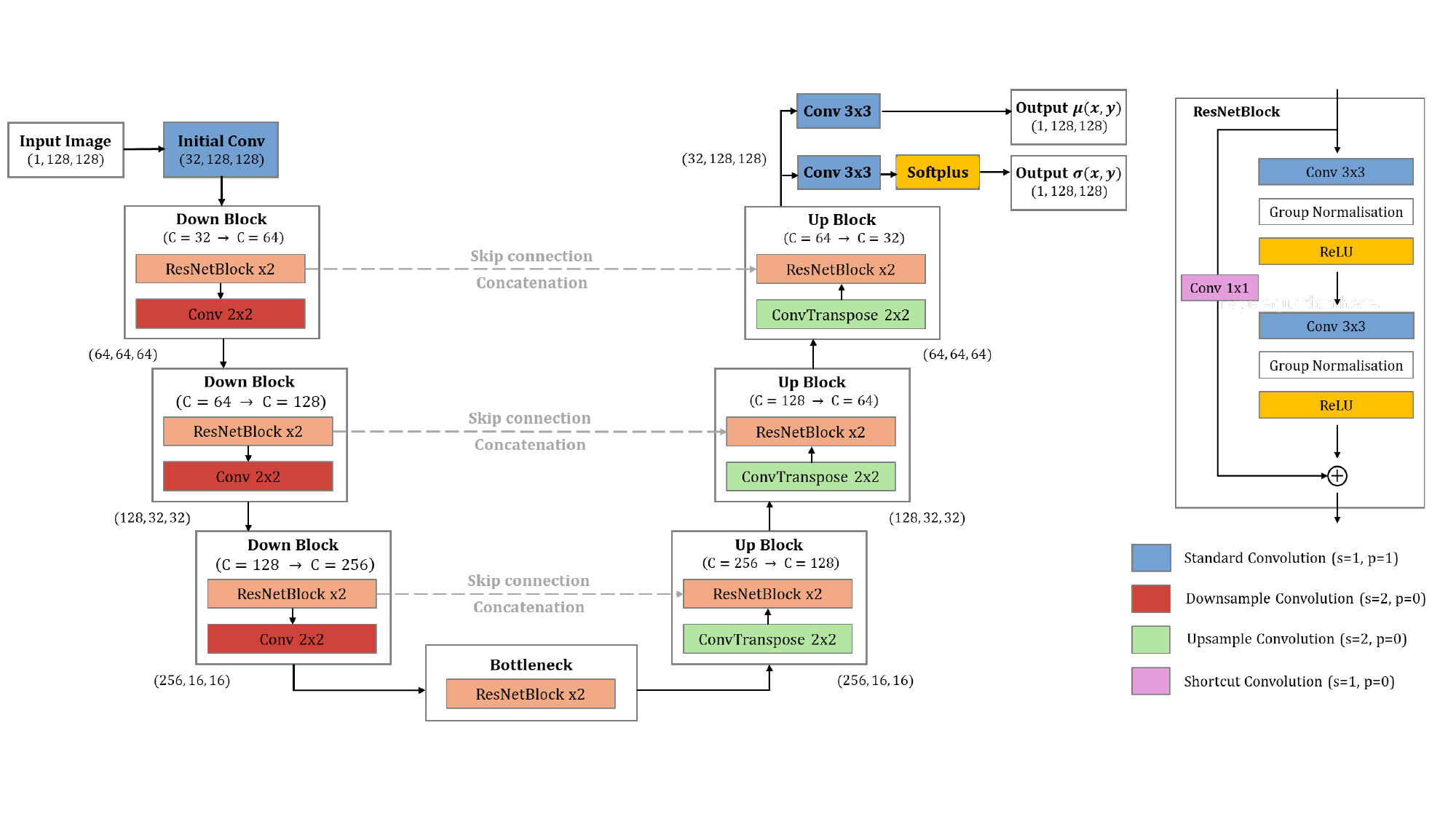}
\caption{Residual U-Net architecture used to learn permeability statistics from mechanical aperture patches. 
The encoder extracts multi-scale geometrical features through residual convolutional blocks with 
progressively increasing channel depth (32–256), while the decoder reconstructs full-resolution 
fields using symmetric upsampling and skip connections. Residual shortcuts are applied within each 
block to improve gradient flow and preserve geometric structure. The output consists of two channels 
corresponding to the log-normal permeability parameters $\mu(x,y)$ and $\sigma(x,y)$.
}
\label{fig:unet_architecture}
\end{figure}

To improve gradient flow and stabilise the optimisation, residual blocks are used throughout the encoder and decoder. Residual learning \cite{ResNet_2016} allows the network to learn perturbations from the identity rather than entire non-linear transformation by using shortcut connections (see Figure~\ref{fig:unet_architecture}). This helps mitigating vanishing-gradient issues and accelerating the convergence, while preserving geometric sharpness \cite{ResNet_vanish_grad, yousef_u-net-based_2023}. The U-Net architecture together with residual formulation thus provide a multi-scale representation well aligned with the physics of rough fractures, allowing the model to learn spatially coherent log-normal statistics $\big(\mu(x,y),\sigma(x,y)\big)$.

The network takes as input a single-channel aperture patch and outputs two channels representing the log-normal permeability parameters at the same resolution, with a Softplus activation enforcing $\sigma(x,y)>0$. Feature depth increases from 32 to 256 channels across three encoder stages, each implemented as residual convolutional blocks followed by $2{\times}2$ strided convolutions for downsampling, used here as a learnable alternative to max pooling. The decoder mirrors this structure by progressively upsampling feature maps using transposed convolutions with skip connections, concatenating the corresponding encoder features at each stage. The entire network architecture is illustrated in Figure~\ref{fig:unet_architecture} and the source code is available at Perez (2026a) \cite{Perez2026_Code}.  

Training is performed using a multi-output loss function that combines data misfit, edge-aware regularisation, and adaptive task weighting. Let $\big(\widehat{\mu}(x,y),\,\widehat{\sigma}(x,y)\big)$ denote the U-Net outputs, the data-fitting terms are defined using the Mean Squared Errors (MSE):
\begin{equation*}
\mathcal{L}_{\mu} = \mathrm{MSE}\,\big(\widehat{\mu}(x,y),\,\mu(x,y)\big),
\quad \text{and} \quad
\mathcal{L}_{\sigma} = \mathrm{MSE}\,\big(\widehat{\sigma}(x,y),\,\sigma(x,y)\big).
\end{equation*}
To promote spatial consistency of permeability structures and better capture high-frequency features, we include an edge-aware regularisation based on Sobel gradients \cite{gou_gradient_2019, Edge_loss_2022, U_Net_Edge_2025}. Let $\nabla_s(\cdot)$ denote the Sobel gradient operator and $\|\cdot\|$ its Euclidean magnitude, the associated loss terms include
\begin{equation*}
\mathcal{L}_{\mu}^{\nabla} = \mathrm{MSE}\,\big(\|\nabla_s \widehat{\mu}(x,y)\|,\, \|\nabla_s \mu(x,y)\|\big),
\quad \text{and} \quad
\mathcal{L}_{\sigma}^{\nabla} = \mathrm{MSE}\,\big(\|\nabla_s \widehat{\sigma}(x,y)\|,\, \|\nabla_s \sigma(x,y)\|\big)
\end{equation*}
where $\nabla_s(\cdot)$ is computed via discrete convolution with the standard Sobel kernels. This formulation penalises discrepancies in spatial derivatives rather than raw values alone, helping the network reproduce sharp permeability transitions in natural rough-walled fractures, while reducing sensitivity to local measurement noise.

We employ GradNorm to adaptively balance the prediction tasks involved in learning $\mu(x,y)$ and $\sigma(x,y)$, eliminating the need to manually tune their relative contributions during the optimisation \cite{Chen2017GradNorm}. Denoting the learnable task weights by $(w_\mu, w_\sigma)$ and the edge regularisation coefficient by $\lambda$, the multi-objective loss is then defined as: 
\begin{equation}
    \mathcal{L} = w_\mu \mathcal{L}_{\mu} + w_\sigma \mathcal{L}_{\sigma} + \lambda\big(w_\mu \mathcal{L}_{\mu}^{\nabla} + w_\sigma \mathcal{L}_{\sigma}^{\nabla}\big),
\label{eq:loss_function}
\end{equation}
where $(w_\mu,w_\sigma)$ are dynamically updated to maintain balanced gradient magnitudes across the two outputs fields, according to the GradNorm criterion. The Sobel-based gradient losses share the same task weights as the data-fidelity losses, $\mathcal{L}_mu$ and $\mathcal{L}_\sigma$, to preserve consistent regularisation across outputs, but are excluded from the GradNorm objective to avoid biasing the task-balancing mechanism toward high-frequency components. For stability and computational efficiency, GradNorm is applied only to the U-Net output layer and the weights are updated every five optimisation steps.

From the 560 paired patches described in Section~\ref{subsec:training_dataset}, the training and validation sets are constructed by splitting the 140 distinct fracture realisations prior to data augmentation, yielding 448 training and 112 validation patches, respectively. All patches derived from a given fracture realisation are assigned exclusively to either the training or the validation set. This prevents leakage of spatial information arising from augmented or spatially correlated representations of the same fracture. We employ the AdamW optimiser with an initial learning rate of $10^{-3}$ and a weight decay of $10^{-2}$, applied to both the network parameters and the learnable loss weights. A cosine annealing schedule is used for learning-rate decay, gradually reducing the step size to $10^{-6}$ over the maximum number of epochs. The model is trained on a Tesla P100-PCIE with 16GB 
of memory for a total of 1000 epochs, using a batch size of 32. The performance of this training strategy, together with the corresponding validation results, is presented in 
Section~\ref{subsec:Results_UNet}.

\subsection{From local statistics to probabilistic permeability fields}
\label{subsec:prob_maps}

Once trained, the Residual U-Net provides a fast surrogate for estimating the spatially varying permeability distribution of a fracture system directly from its mechanical aperture field. Given a high-resolution aperture map $a_m(x,y)$, the local permeability distribution is reconstructed on a pixel basis as
\begin{equation}
    K(x,y) \sim \mathrm{Log}\,\mathcal{N}\big(\widehat{\mu}(x,y),\,\widehat{\sigma}^2(x,y)\big),  \qquad \forall (x,y)\in\OmegaD{2}.
    \label{eq:UNet_perm_field}
\end{equation}
This transformation yields, for each spatial location, not a single permeability value but a continuous distribution encoding the expected hydraulic response and its associated uncertainty. Indeed, from the predicted parameters $\big(\widehat{\mu}(x,y),\widehat{\sigma}(x,y)\big)$, we derive key probabilistic descriptors of the permeability field, namely the expected value (mean behaviour), the most probable outcome (mode) and quantile bounds representing local uncertainties: 
\begin{subequations}
\label{eq:K_descriptors}
    \begin{align}
        \mathbb{E}[K(x,y)] &= \exp\!\left(\widehat{\mu}(x,y) + \frac{1}{2}\widehat{\sigma}^2(x,y)\right)\label{eq:K_descriptors:a} \\[2mm]
        \mathrm{Mode}[K(x,y)] &= \exp\!\left(\widehat{\mu}(x,y) - \widehat{\sigma}^2(x,y)\right)\label{eq:K_descriptors:b} \\[2mm]
        Q_p[K(x,y)] &= \exp\!\left(\widehat{\mu}(x,y) + \phi_{p}\,\widehat{\sigma}(x,y)\right), \qquad p\in[0,1]\label{eq:K_descriptors:c}
    \end{align}
\end{subequations}
where $\phi_{p}$ denotes the p-quantile of the standard normal distribution. In practice, we compute the 95\% confidence interval using the quantiles $Q_{0.025}$ and $Q_{0.975}$ as lower and upper bounds on the permeability maps, respectively. Together, these fields allow for direct visualisation of flow-relevant structures (\emph{e.g.}, channelling pathways, contact areas and flow barriers), while simultaneously illustrating how uncertainty influences or amplifies them. This provides a probabilistic alternative to deterministic aperture-permeability maps derived from Cubic-Law formulations. 

Given its fully convolutional design (see Section~\ref{subsec:UNET}), the network can be applied to fracture domains of arbitrary size and geometry, making it naturally suitable for upscaling. A single forward evaluation produces the complete set of permeability descriptors \eqref{eq:K_descriptors} across the entire domain, providing full probabilistic estimates rather than a single deterministic field (\emph{e.g.}, obtained via Direct Numerical Simulations). Because the inference requires only one forward pass, probabilistic permeability maps can be computed for high-resolution natural fractures at negligible cost relative to Stokes-based inversion. This efficiency enables large-scale Monte-Carlo exploration and renders uncertainty quantification tractable for complex fracture systems and, ultimately, for fracture networks. 

In summary, starting from uncertain mechanical aperture measurements, our framework produces probabilistic permeability fields that are physically consistent, capture local heterogeneity and uncertainty, and are scalable to realistic geological geometries. Section~\ref{subsec:Results_Kmap} presents the probabilistic permeability maps obtained for
the natural fracture geometries introduced in Section~\ref{sec:Geological_frac}.

\subsection{Darcy flow-based upscaling of the uncertainties}
\label{subsec:Darcy_upscaling}

The permeability fields predicted by the U-Net provide a spatially resolved probability distribution of $K(x,y)$ across the fracture plane (see Section~\ref{subsec:prob_maps} and \ref{subsec:Results_Kmap}). To quantify how these local uncertainties propagate into large-scale hydraulic responses, we compute upscaled permeability distributions using steady-state Darcy flow. All the flow-based upscaling simulations are conducted using the MATLAB Reservoir Simulation Toolbox (MRST) \cite{Lie_2019}.

For a deterministic permeability field $k(x,y)$ defined over the fracture domain $\Omega^{2D}$, Darcy’s law for incompressible flow reads
\begin{equation}
    \mathbf{v}(x,y) = - \frac{k(x,y)}{\mu} \nabla p,
    \qquad 
    \nabla \cdot \mathbf{v} = 0,
\end{equation}
where $\mathbf{v}$ denotes the Darcy velocity (in $m.s^{-1}$), $\mu$ is the dynamic viscosity (in $Pa.s$) and $p$ the pressure field (in $Pa$). A constant pressure drop is applied between the inlet and outlet boundaries, while no-flow conditions are prescribed along the remaining boundaries. The upscaled permeability is obtained from the average outflow velocity $q$ and the pressure drop $\Delta p$ over the domain length $L$ (taken in the flow direction):
\begin{equation}
   K_{\mathrm{eff}} = \frac{\mu q L}{\Delta p},
   \label{eq:Keff_Darcy}
\end{equation}
where $q$ is computed as the ratio of the volumetric discharge trough the outlet $Q$ (in $m^3.s^{-1}$) by the cross sectional area $A$ (in $m^2$). 

To propagate local permeability uncertainties to the scale of geological fracture systems, we perform Darcy flow-based upscaling directly on the four probabilistic descriptors generated by the U-Net: the lower quantile, expected value, mode, and upper quantile of the log-normal permeability field (see Eq. \eqref{eq:K_descriptors}). Each map represents a physically plausible realisation of the heterogeneous permeability under uncertain conditions, while preserving the spatial structures inherited from the aperture distribution which is essential for obtaining physically meaningful upscaled uncertainty ranges. 

A Darcy simulation is performed for each descriptor field, and the corresponding effective permeabilities are computed through Eq. \eqref{eq:Keff_Darcy}. The resulting four upscaled permeability values
\begin{equation}
\big\{ K_{Q_{0.025}},\;
       K_{\mathrm{mode}},\;
       K_{\mathrm{mean}},\;
       K_{Q_{0.975}} \big\}  
\label{eq:upscaled_perm_ranges}
\end{equation}
provide a direct quantification of how local uncertainty in aperture-permeability mapping translates into probabilistic hydraulic responses at the fracture scale. To represent this uncertainty range in compact form, we approximate $K_{\mathrm{eff}}$ by a log-normal distribution whose parameters are determined from the upscaled mean and mode, while the lower and upper quantiles constrain the confidence intervals. This yields an analytical probability distribution for $K_{\mathrm{eff}}$ that remains consistent with the local uncertainties inferred by the U-Net, captures the non-linear dependence of effective permeability on spatially heterogeneous $k(x,y)$ during Darcy-based upscaling, and reflects the connectivity of high-permeability pathways across the fracture system. 

To assess the validity of this analytical approximation, we further perform a Monte Carlo propagation of uncertainty based on the predicted log-normal parameters. Uncertainty propagation is introduced through a structure-preserving latent perturbation represented by a fully correlated Gaussian variable $Z \sim \mathcal{N}(0,1)$. Permeability realisations are therefore generated as
\begin{equation}
K^{(n)}(x,y) = \exp\!\left(\widehat{\mu}(x,y) + \widehat{\sigma}(x,y)\,Z_n\right), \qquad Z_n \overset{\mathrm{iid}}{\sim} \mathcal{N}(0,1)
\label{eq:perm_real_MC}
\end{equation}
which preserves the spatial organisation of the permeability field while modulating local hydraulic efficiency according to the predicted uncertainty amplitude. Darcy upscaling is performed for each realisation to obtain a Monte Carlo distribution of $K_{\mathrm{eff}}$, which is then compared to the analytical log-normal representation. In this formulation, $Z$ represents a coherent, globally correlated uncertainty in the aperture–permeability mapping, i.e. a mapping-level perturbation, while preserving the dominant flow topology inherited from the observed aperture geometry.
Introducing instead a correlated Gaussian random field would correspond to a different uncertainty model, as it would inject additional spatial variability not constrained by the aperture field, i.e. stochastic geometry variability, and is therefore outside the scope of the present work. 

In summary, the proposed upscaling procedure translates pixel-scale uncertainty into a probabilistic estimate of fracture-scale permeability, grounded in both aperture geometry and flow physics. Section~\ref{subsec:Results_Darcy} presents the resulting $K_{\mathrm{eff}}$ distributions for the natural fracture datasets introduced in Section~\ref{sec:Geological_frac} and compares them with Cubic-Law approximations and reference Stokes-flow simulations.

\section{Geological fracture dataset}
\label{sec:Geological_frac}

We evaluate our probabilistic upscaling framework on two fracture datasets: (i) some natural shear fractures obtained via high-resolution micro-tomography, and (ii) their synthetic counterparts designed to reproduce similar characteristics while offering simplified geometries highlighting the dominant channels. The natural samples provide representative cases of geological complexity, whereas the synthetic fractures serve as simplified benchmarks for validating the hydraulic upscaling and uncertainty propagation.

Here, we present the two datasets, describe how aperture fields are extracted from the tomographic volumes, and compare the fracture statistics and permeability estimates using Cubic-Law and Stokes-flow models. These comparisons will later provide the reference baseline for evaluating the reliability of our probabilistic upscaling results in Section~\ref{sec:Results}.

\subsection{Natural shear fracture dataset}
\label{subsec:natural_real}

The natural fractures analysed in this study were extracted from core material recovered within the damage zone of the Little Grand Wash Fault near Green River Basin, Utah \cite{KAMPMAN_2014_22, KAMPMAN_2014_51}. The fractured sample plugs originate from the Jurassic Carmel Formation, representing a shallow-marine succession of claystone, siltstone and evaporite layers, acting as a caprock for the underlying Navajo Formation. Several fractures in the samples exhibit clear indicators of shear displacement, including mismatched surface geometries and locally interlocked asperities. Such features provide strong evidence that the fractures formed in situ during fault activity. Two of these natural shear fractures are selected here to evaluate the performance of the proposed uncertainty-aware upscaling framework.

Synchrotron-based, high-resolution X-ray micro-tomography ($\mu$CT), with a voxel size of $2.75\mu m$, was performed to capture the full 3D geometry of the void space. The tomographic volumes were processed to retain only the fracture and its immediate surrounding rock matrix, yielding datasets of size $1480 \times 1873 \times 457$ voxels and $1480 \times 1775 \times 467$ voxels for the two natural fractures, respectively. These resolutions are sufficient to resolve the geological roughness, contact zones, and channelised, branching void structures characteristic of shear displacement in natural fractures. Further details on the core samples and imaging procedure are provided in Phillips et al. (2024) \cite{Phillips2024}.

\begin{figure}[t!]
\centering
\raisebox{0.25\height}{
\begin{subfigure}{.49\textwidth}
    \caption{}
    \centering
    \includegraphics[width=\linewidth]{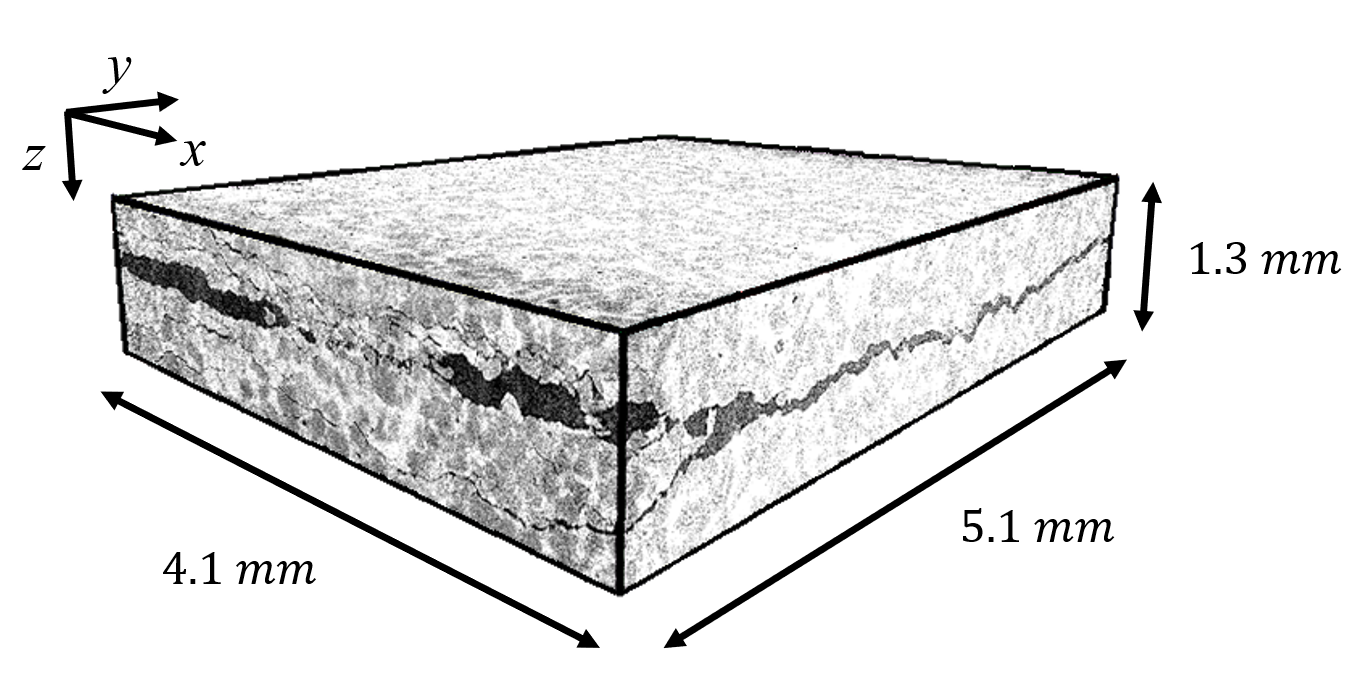}
    \label{fig:3D_vol}
\end{subfigure}
}%
\hspace{-1mm}
\begin{subfigure}{.49\textwidth}
    \caption{}
    \centering
    \includegraphics[width=\linewidth]{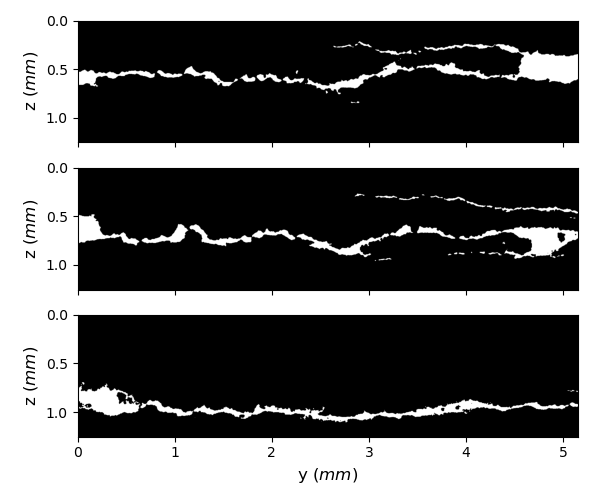}
    \label{fig:cross_sec}
\end{subfigure}
\caption{(a) Greyscale $\mu$CT volume rendering of one natural shear fracture, showing physical dimensions and axis orientation ($z$ denoting the fracture depth). (b) Representative segmented slices extracted along the $y$-axis after image processing. These cross-sections highlight the complex aperture structure, including contact zones, locally disconnected void clusters and branching pathways, that influence the hydraulic behaviour of the fracture.}
\label{fig:natural_fracture_geometry}
\end{figure}

Image segmentation is performed using a three-dimensional anisotropic diffusion filter, which reduces noise while preserving geometrical edges, followed by histogram-based thresholding. A connectivity check is then used to eliminate isolated artefacts. The resulting binary images reveal interlocked asperities and locally disconnected void clusters that complicate the geometry, producing mechanical aperture fields that deviate strongly from classical cubic-law assumptions. Aperture thickness fluctuates over micrometre scales, and multiple potential flow pathways coexist, opening or closing depending on subtle mismatches between the opposing walls. This combination of roughness and multivalued aperture structure generates strong spatial variability in hydraulic response, making this dataset a challenging test case for uncertainty-aware permeability upscaling. Figure~\ref{fig:natural_fracture_geometry} displays the greyscale $\mu$CT volume together with several segmented cross-sections, highlighting the roughness and multivalued aperture geometry of one of the natural shear fractures.

\subsection{Simplified synthetic fractures for hydraulic validation}
\label{subsec:synthetic}

To provide a reference case alongside the natural shear fractures described in Section~\ref{subsec:natural_real}, we construct simplified synthetic fractures derived from the same $\mu$CT dataset. The synthetic fractures preserve the large-scale geometry and aperture variability of the natural fracture (see Figures~\ref{fig:Aperture_maps_comp}c and \ref{fig:synthetic_comp}), while simplifying the topologically complex features that complicate the interpretation of local aperture and flow connectivity.

The simplified synthetic geometries are built by collapsing the full 3D void structure into a single connected band. For each in-plane location $(x,y)$, the local fracture thickness and its mid-position along the $z$-direction are extracted from the segmented volume. These two fields, representing the amount of open space and its average location, are then used to reconstruct a continuous synthetic fracture geometry. This procedure retains the overall morphology of the natural fracture, highlighting the dominant channel, but removes disconnected clusters and branch bifurcations.

The resulting synthetic fractures provide simplified yet realistic counterparts, enabling a clean comparison with the original natural fractures. In addition, its hydraulically interpretable geometry admits a relatively simple definition of a single local aperture field, rendering it an effective benchmark for evaluating our probabilistic permeability mapping and upscaling approach.

\subsection{Thickness-weighted aperture averaging}
\label{subsec:aperture_weight}

For the two datasets introduced in Sections~\ref{subsec:natural_real} and~\ref{subsec:synthetic}, we need to establish a unified and consistent way to estimate the mechanical aperture field $a_m(x,y)$ of complex 3D fracture geometries. Indeed, the segmented $\mu$CT volumes of geological fractures contain, at each in-plane location $(x,y)$, a potential stack of open and closed voxels along the fracture-normal direction $z$ (see Figure~\ref{fig:cross_sec}). Several branches may coexist within the same column, so a single in-plane location in the natural fracture can be associated with multiple aperture measurements, whereas the synthetic fracture provides a single aperture value at each location. Similar ambiguities in defining a representative mechanical aperture have also been highlighted for rough and partially cemented fractures by Landry et al. (2024) \cite{Landry_2024}, who demonstrated the effects of different image-based aperture measurement methods on the estimated hydraulic properties. 

\begin{figure}[t!]
    \centering
    \begin{tikzpicture}
        \node[anchor=south west] at (0,0)
            {\includegraphics[width=\linewidth]{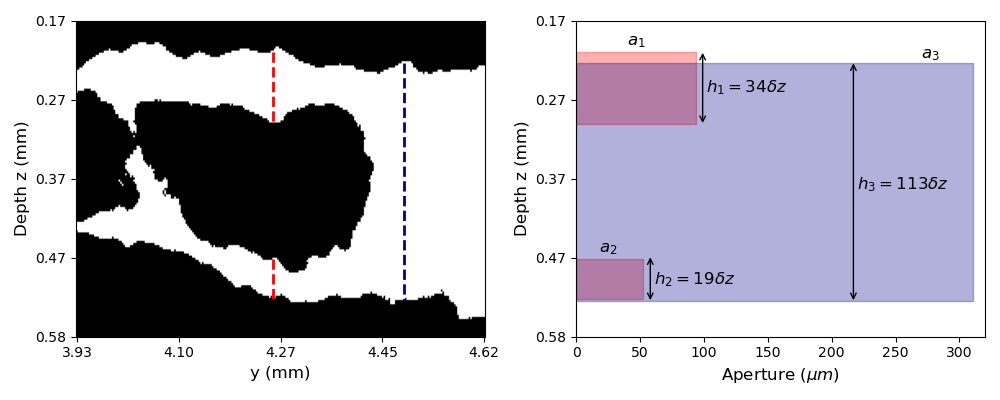}};
        \node at (0.04\linewidth,0.44\linewidth) {\small (a)};
        \node at (0.54\linewidth,0.44\linewidth) {\small (b)};
    \end{tikzpicture}
    \caption{(a) Binary cross-section of a segmented $\mu$CT fracture (zoomed), with red and blue dashed lines marking two columns used to illustrate the computation of the thickness-weighted vertical aperture. (b) Schematic representation of the aperture profiles extracted along these columns. For the red line, two branches coexist with apertures $a_1,\,a_2$ and heights $h_1,\,h_2$, so that Eq. \eqref{eq:thickness_weighted_aperture} gives $a_m(x,y) = (a_1 h_1 + a_2 h_2)/(h_1 + h_2)$. For the blue line, which corresponds to a single branch, equation \eqref{eq:thickness_weighted_aperture} yields $a_m(x,y)=a_3$.} 
\label{fig:thickness_weighted}
\end{figure}

To obtain a mechanically interpretable aperture field for both fracture types, we introduce the \emph{thickness-weighted mean} of the vertical aperture (see Figure~\ref{fig:thickness_weighted}). Let $\mathcal{O}(x,y) = \{\, z: a(x,y,z)>0 \,\}$ denote the set of depths $z$ at which the fracture is locally open. For each open voxel, $a(x,y,z)$ represents the local 3D aperture \emph{i.e.}, the normal separation between the two opposing walls. We then define the  thickness-weighted mean aperture as
\begin{equation}
    a_m(x,y)
    = \frac{\displaystyle \sum_{z\in O(x,y)} a(x,y,z)\,\delta z}
           {\displaystyle \sum_{z\in O(x,y)} \delta z}
    \label{eq:thickness_weighted_aperture}
\end{equation}
where $\delta z$ is the voxel size in the fracture-normal direction.

This definition offers several advantages. First, it yields a geometric aperture measured only where the fracture is actually open, and therefore reduces to the standard vertical aperture for synthetic fractures that contain a single connected channel. Second, unlike the maximum-aperture criterion applied to partially cemented fractures in Landry et al. (2024) \cite{Landry_2024}, it accounts for the contribution of all coexisting branches to the hydraulic behaviour. By weighting each branch by its local open thickness, the definition naturally blends their influence in a manner consistent with their volumetric importance. In fact, \eqref{eq:thickness_weighted_aperture} corresponds to the expected gap encountered when sampling a point uniformly within the open sub-volume, thereby providing a probabilistic interpretation of the local mechanical aperture in complex fracture geometries.

\begin{figure}[t!]
    \centering
    \begin{tikzpicture}
        \node[anchor=south west] at (0,0)
            {\includegraphics[width=\linewidth]{Figs/Aperture_maps_comp.png}};
        \node at (0.04\linewidth,0.35\linewidth) {\small (a)};
        \node at (0.25\linewidth,0.35\linewidth) {\small (b)};
        \node at (0.46\linewidth,0.35\linewidth) {\small (c)};
        \node at (0.75\linewidth,0.35\linewidth) {\small (d)};
    \end{tikzpicture}
   \caption{Comparison of mechanical aperture fields of fracture \#1: (a) Thickness-weighted mean aperture defined by Eq.~\eqref{eq:thickness_weighted_aperture}, which blends coexisting branches according to their open thickness.  (b) Maximum vertical aperture extracted from the real fracture, highlighting dominant wide channels. (c) Aperture field of the corresponding synthetic fracture, exhibiting a single connected band. (d) Relative aperture difference between the thickness-weighted mean and maximum aperture. The comparison illustrates the spatial variability induced by different aperture definitions and motivates the probabilistic treatment adopted in this work.}
\label{fig:Aperture_maps_comp}
\end{figure}
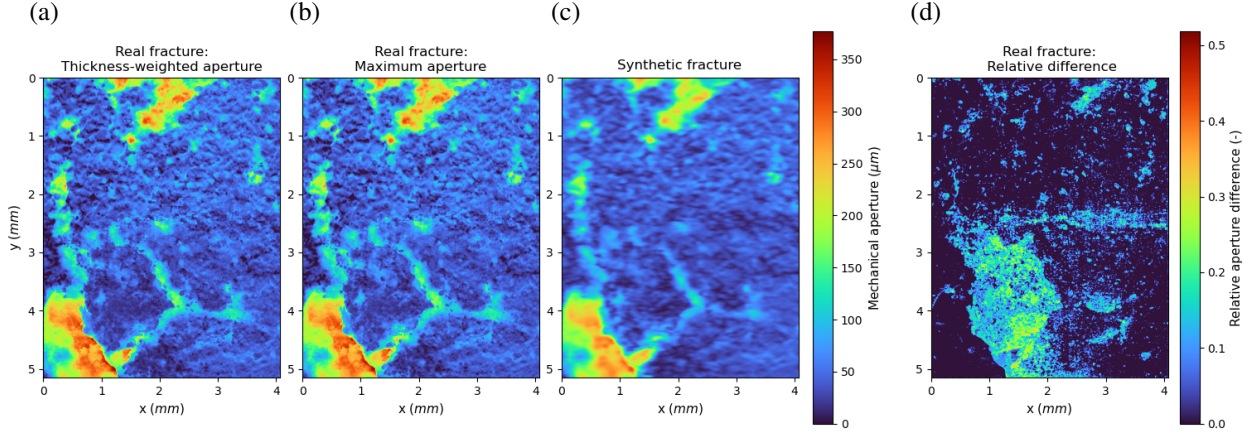

Complementary descriptors, such as the maximum aperture, may also be retained to highlight dominant wide channels, and the corresponding aperture fields are compared in Figure~\ref{fig:Aperture_maps_comp}. In this work, the thickness-weighted mean \eqref{eq:thickness_weighted_aperture} is therefore not interpreted as a unique or exact aperture, but as a consistent geometric and probabilistic descriptor derived from the $\mu$CT data. It is important to emphasise that the mechanical aperture field defined in Eq.~\ref{eq:thickness_weighted_aperture} is not, by itself, hydraulically interpretable. Direct use of $a_m(x,y)$ in two-dimensional flow models, for instance through Cubic-Law formulations, implicitly assumes a direct aperture-permeability correspondence and is shown to systematically overestimate fracture permeability for rough and multivalued aperture structures. In the present framework, $a_m(x,y)$ instead serves as the uncertain geometric input to our hybrid probabilistic upscaling that learns an effective in-plane permeability field $K(x,y)$ consistent with the hydraulic response of the full 3D fracture geometry. This mapping defines a quasi 2D hydraulic representation, with uncertainty reflecting the limitations inherent to this dimensional reduction. Discrepancies arising from alternative aperture definitions are, in particular, treated as part of the measurement uncertainty on $a_m(x,y)$ and are explicitly incorporated into the Bayesian formulation introduced in Section~\ref{subsec:Bayesian_inference}.

\subsection{Aperture maps, hydraulic responses and fracture statistics}
\label{subsec:fracture_comparison}

\begin{figure}[t!]
\centering
\raisebox{0.05\height}{
\begin{subfigure}{.49\textwidth}
    \caption{}
    \centering
    \includegraphics[width=\linewidth]{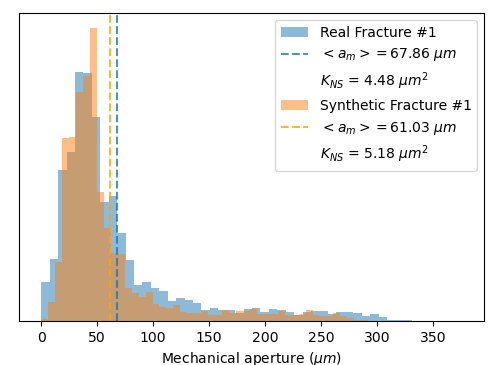}
    \label{fig:synthetic_dist}
\end{subfigure}
}%
\hspace{-1mm}
\begin{subfigure}{.49\textwidth}
    \caption{}
    \centering
    \includegraphics[width=\linewidth]{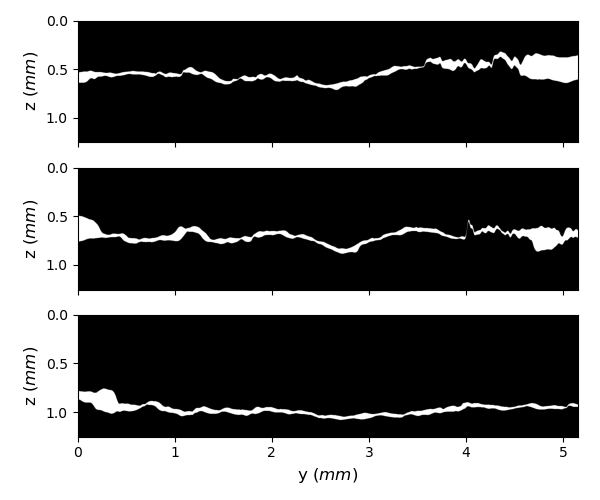}
    \label{fig:cross_sec_synthetic}
\end{subfigure}
\caption{(a) Mechanical aperture distributions for natural shear fracture \#1 and its synthetic counterpart. The mean mechanical aperture, denoted $\langle a_m \rangle$, and the Stokes-based permeability $K_{NS}$ are reported for each configuration. (b) Representative segmented cross-sections extracted along the $y$-axis for the synthetic fracture, taken at the same locations as for the natural fracture.
These cross-sections highlight the simplified aperture geometry of the synthetic fracture, which forms a single connected band compared to the multivalued structure of the natural fracture (see Figure \ref{fig:cross_sec}).}
\label{fig:synthetic_comp}
\end{figure}

The geometric and hydraulic characteristics of the natural shear fractures and their synthetic counterparts are synthesised through a joint analysis of mechanical aperture maps, spatial statistics, and permeability estimates. For consistency, the notations adopted here for the different permeability measures follow those of Perez et al. (2025) \cite{GRL_Perez_2025}: $K_{CL}$ denotes the permeability estimated from classical Cubic Law formulation, $K^{\, a_m}_{D}$ the Darcy-upscaled permeability based on local Cubic-Law assumption, and $K_{NS}$ the Stokes permeability obtained from Direct Numerical Simulations. The reader is referred to Perez et al. (2025) \cite{GRL_Perez_2025} for a detailed description of the corresponding modelling assumptions and numerical implementations.

Figure~\ref{fig:synthetic_dist} compares the mechanical aperture distributions of the natural shear fracture \#1 and its synthetic counterpart, computed from the thickness-weighted mean aperture field $a_m(x,y)$ defined in Eq.~\eqref{eq:thickness_weighted_aperture}. While both datasets exhibit comparable mean mechanical apertures ($\langle a_m \rangle \approx 60$--$70~\mu$m), the natural fracture displays a slightly broader distribution with higher variance, reflecting the multivalued aperture structure induced by roughness, contact zones and branching pathways. These geometric differences translate directly into distinct hydraulic responses. 

Table~\ref{table1} summarises the corresponding geometric and hydraulic properties, including the mean aperture $\langle a_m \rangle$, aperture standard deviation $\sigma_{a_m}$, relative roughness $\varepsilon = \sigma_{a_m} / \langle a_m \rangle$,
correlation length $\xi_{frac}$, and permeability estimates obtained using different modelling approaches. Permeability estimates based on the classical Cubic-Law formulation $K_{CL}$, which assumes a direct local mapping between mean aperture $\langle a_m \rangle$ and transmissivity, systematically overestimate natural fracture permeabilities compared to the Stokes-consistent reference $K_{NS}$, as previously demonstrated on synthetic geometries in Perez et al. (2025) \cite{GRL_Perez_2025}. This bias persists when local Cubic-Law permeabilities are evaluated from the mechanical aperture field and subsequently upscaled through Darcy flow, yielding $K^{\, a_m}_{D}$ (Table~\ref{table1}). Although the latter approach accounts for the spatial heterogeneity and connectivity of the aperture field, it remains constrained by power-law assumption at the local scale and therefore cannot fully capture the impact of roughness-induced flow redistribution and partial closures. These results confirm the limitations of deterministic aperture-permeability mappings when applied to complex natural fracture geometries.
\begin{table}[pb!]
\centering
\begin{tabular}{lcccccc}
\hline \multirow{2}{*}{Features} & \multirow{2}{*}{Notations} & \multirow{2}{*}{Units} &\multicolumn{2}{c}{Fracture \#1} & \multicolumn{2}{c}{Fracture \#2} \\
& & & Real & Synthetic & Real & Synthetic\\
\hline 
Mean mechanical aperture & $\langle a_m \rangle$ & $\mu m$ & 67.86 &  61.03 & 140.5 & 136.6 \\
Aperture standard deviation & $\sigma_{a_m}$ & $\mu m$ & 61.02 & 51.89 & 65.13 & 63.44  \\
Relative roughness: ${\sigma_{a_m}}/{\langle a_m \rangle}$& $\varepsilon $ & / & 0.89 & 0.85 & 0.46 & 0.46 \\
Correlation length & $\xi_{frac}$ & $mm$ & 1.55 & 2.28 &  1.44 & 1.57 \\
Cubic Law (CL) Permeability & $K_{CL}$ & $\mu m^2$ & 383.8 & 310.4 & 1645.4 & 1555.2 \\
Darcy Permeability: Local CL & $K^{\, a_m}_{D}$ & $\mu m^2$ & 210.9 & 188.5 & 1135.3 & 1066.8 \\
Stokes Permeability & $K_{NS}$ & $\mu m^2$ & 4.48 & 5.18 & 62.08 & 75.19  \\
\hline
\end{tabular}
\caption{Comparison of geometric aperture descriptors and permeability estimates for the natural shear fractures and their synthetic counterparts. Reported quantities include mean and standard deviation of the mechanical aperture, relative roughness, correlation length, and effective permeabilities computed using Cubic-Law, Darcy-upscaled local Cubic-Law, and Stokes-consistent approaches. Directional quantities such as the correlation lengths and permeability values are given in the main flow direction, taken as the $x$-direction (with $L_x = 4.1\mathrm{mm}$ for both fractures).}
\label{table1}
\end{table}

\begin{figure}[ht!]
\centering
\begin{subfigure}{.49\textwidth}
    \caption{}
    \centering
    \includegraphics[width=\linewidth]{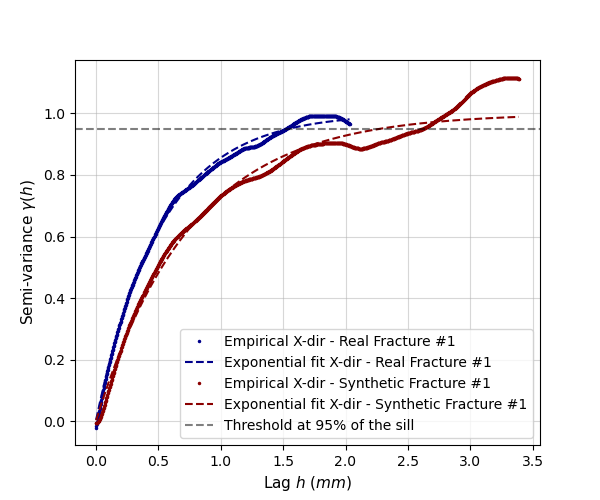}
    \label{fig:CorL_frac1}
\end{subfigure}
\hspace{-1mm}
\begin{subfigure}{.49\textwidth}
    \caption{}
    \centering
    \includegraphics[width=\linewidth]{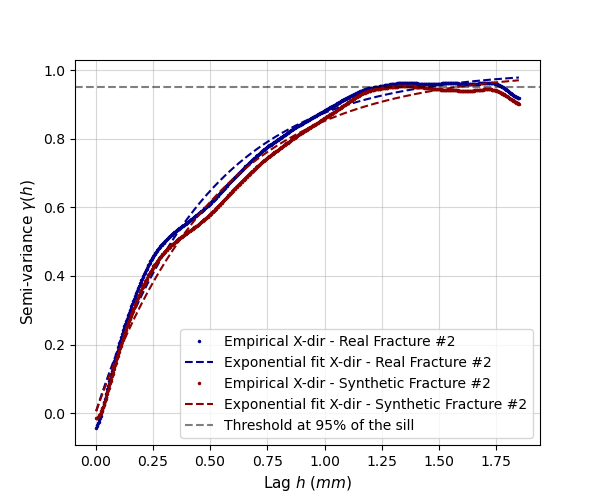}
    \label{fig:CorL_frac2}
\end{subfigure}
\caption{Variogram analysis in the main flow direction, taken as the $x$-direction, used to determine the correlation lengths of both fractures: (a) Real shear fracture \#1 and its synthetic counterpart. (b) Real shear fracture \#2 and its synthetic counterpart. Variograms are fitted with an exponential model. The correlation length is the lag distance $h$ at which the variogram reaches 95\% of the sill, which has first been normalized.}
\label{fig:CorL}
\end{figure}

Figure~\ref{fig:Aperture_maps_comp} illustrates how different aperture definitions modify the spatial organisation of flow-relevant features. The thickness-weighted mean aperture blends coexisting branches according to their volumetric contribution, whereas the maximum aperture isolates dominant wide channels and amplifies local contrasts. The relative difference between these two fields (Figure~\ref{fig:Aperture_maps_comp}d) reveals spatially localised discrepancies, reinforcing the need to treat mechanical aperture as an uncertain geometric descriptor rather than a unique hydraulic proxy.

Beyond marginal aperture statistics, the spatial organisation of aperture fluctuations plays a critical role in controlling hydraulic behaviour. To quantify this, experimental variograms of the thickness-weighted aperture are computed along the main flow direction ($x$-direction) for both the natural fractures and their synthetic counterparts (see Figure~\ref{fig:CorL}). The variograms are fitted with exponential models, and the correlation length $\xi_{frac}$ is defined as the lag distance at which the variogram reaches $95\%$ of its sill. As reported in Table~\ref{table1}, the synthetic
fractures exhibit longer correlation lengths than their natural counterparts, reflecting the smoother and more continuous aperture structure induced by geometric simplification. For fracture \#2, however, the difference between natural and synthetic cases is markedly reduced. The variograms display relatively similar correlation lengths, consistent with the cross-sectional observations (see Figure~\ref{fig:cross_sec_NF2} in \ref{app:NF2}), which show a less pronounced branching structure compared to fracture \#1. This reduced topological complexity results in comparable spatial organisation of aperture fluctuations in both the natural and synthetic representations of fracture \#2 (see Figure~\ref{fig:real_vs_synthetic_apertures_NF2} in \ref{app:NF2}).

Together, these statistics provide a compact characterisation of fracture heterogeneity and establish the geometric and hydraulic basis for validating our probabilistic permeability mapping and upscaling framework. 

\section{Results}
\label{sec:Results}

This section presents the results of the proposed probabilistic upscaling framework developed in Section~\ref{sec:Up_fram}. We first validate the U-Net-based learning stage and assess its ability to generate probabilistic permeability maps on the large-scale natural fracture geometries introduced in Section~\ref{sec:Geological_frac}. We then analyse how local permeability uncertainties propagate to the scale of the effective hydraulic response.

\subsection{Local training and validation of the U-Net}
\label{subsec:Results_UNet}

The performance of the U-Net model (Figure~\ref{fig:unet_architecture}) in learning the probabilistic mapping from mechanical aperture fields to local permeability statistics is first assessed at the patch scale. The network performances are evaluated on the training and validation datasets described in Section~\ref{subsec:training_dataset} and \ref{subsec:UNET}.

\begin{figure}[t!]
    \centering
    \begin{tikzpicture}
        \node[anchor=north west] at (0,0)
            {\includegraphics[width=\linewidth]{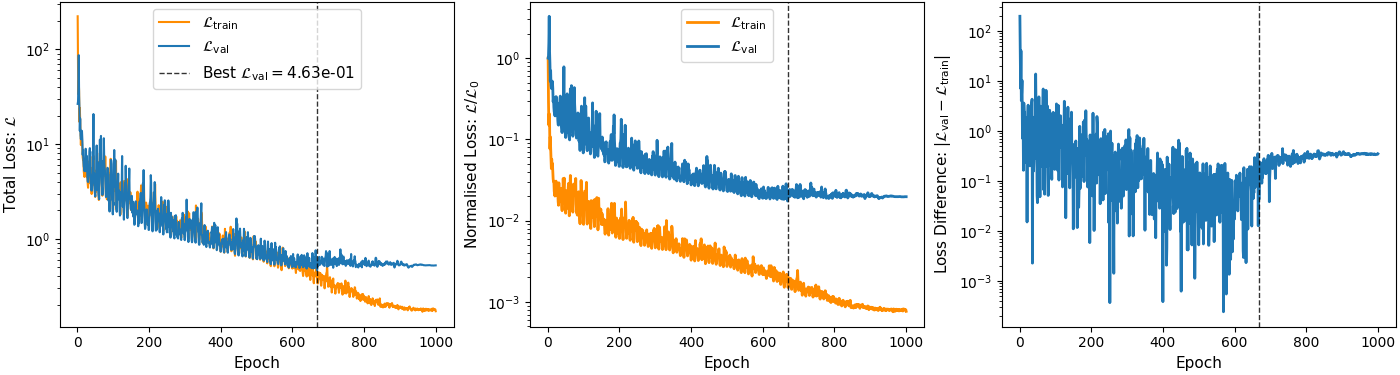}};
        \node at (0.05\linewidth,0.03\linewidth) {\small (a)};
        \node at (0.38\linewidth,0.03\linewidth) {\small (b)};
        \node at (0.72\linewidth,0.03\linewidth) {\small (c)};
    \end{tikzpicture}
   \caption{
    Training history of the Residual U-Net: (a) Absolute training and validation losses, $\mathcal{L}_{\mathrm{train}}$ and $\mathcal{L}_{\mathrm{val}}$, as functions of the number of epochs.The epoch corresponding to the minimum validation loss is indicated by the vertical dashed line. (b) Normalised losses, obtained by scaling each curve by its initial value, highlighting the relative convergence rate of the optimisation. (c) Generalisation gap magnitude, defined as the absolute difference between training and validation losses, used to assess overfitting and training stability. Both absolute and normalised losses decrease jointly during training, indicating effective optimisation and convergence. Near the optimal epoch, a saturation of validation performance is observed, leading to a moderate increase in the generalisation gap. 
    }
\label{fig:Loss_conv}
\end{figure}
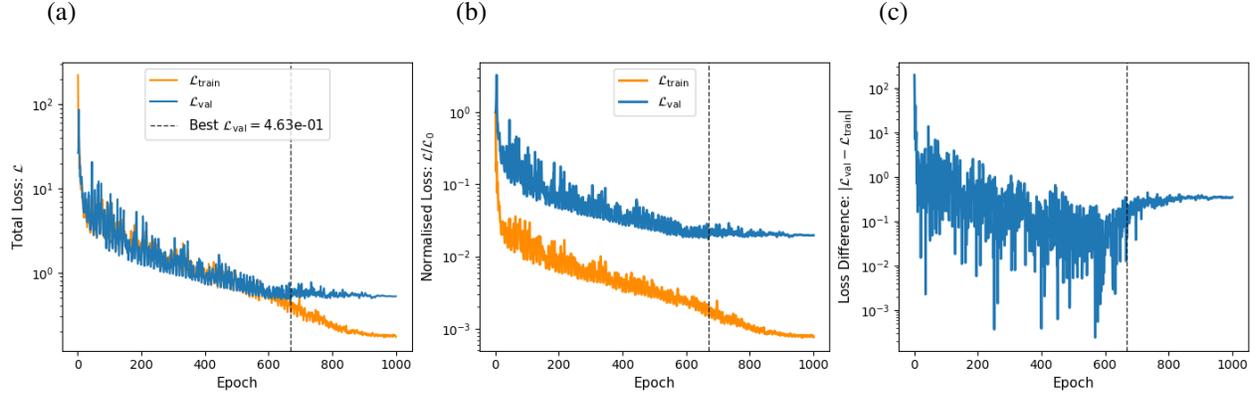

Figure~\ref{fig:Loss_conv} summarises the training history over the total 1000 epochs. The training and validation losses, denoted $\mathcal{L}_{train}$ and $\mathcal{L}_{val}$, decrease jointly and progressively stabilise as training progresses, with the minimum validation loss reached at epoch 668. This behaviour indicates stable convergence without evidence of early overfitting, followed by a saturation of validation performance beyond the optimal epoch. The normalised loss curves (Figure~\ref{fig:Loss_conv}b), obtained by scaling each loss by its initial value, exhibit similar decreasing trend for the training and validation datasets with effective learning during the early stages of optimisation. The generalisation gap magnitude (Figure~\ref{fig:Loss_conv}c), defined as the absolute difference between training and validation losses, remains limited throughout training decreasing during the main convergence phase and increasing moderately beyond the optimal epoch. This behaviour confirms the stability of the learning process and the controlled generalisation capability of the network. These generalisation properties are further examined in Section~\ref{subsec:generalisation} when applying the framework to the natural fracture geometries.

\begin{figure}[ht!]
    \centering
    \includegraphics[width=0.8\linewidth]{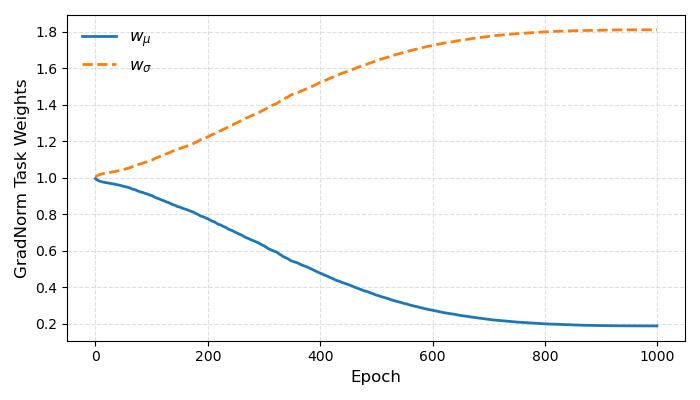}
    \caption{Evolution of the GradNorm task weights during training for the regression objectives associated with the log-normal permeability parameters $\mu(x,y)$ and $\sigma(x,y)$. The adaptive weights $w_\mu$ and $w_\sigma$ are adjusted to balance the gradient magnitudes of the tasks at the U-Net output layer. The gradual decrease of $w_\mu$ and increase of $w_\sigma$ indicate a dynamic rebalancing of the learning effort between the two outputs, enabling stable joint optimisation without manual tuning of loss weights.}
\label{fig:Weights_evol}
\end{figure}

Figure~\ref{fig:Weights_evol} shows the evolution of the adaptive task weights $(w_\mu, w_\sigma)$ associated with the regression of the log-normal permeability parameters $\mu(x,y)$ and $\sigma(x,y)$ (see Eq.~\eqref{eq:loss_function}). As described in Section~\ref{subsec:UNET}, these weights are dynamically updated during training using the GradNorm strategy (see~\ref{app:gradnorm} for details). The gradual decrease of $w_\mu$ and corresponding increase of $w_\sigma$ reflect an automatic rebalancing of the learning effort as training progresses.  Early optimisation stages prioritise the prediction of $\mu(x,y)$, which captures the dominant spatial organisation of permeability. As training proceeds, increased emphasis is placed on $\sigma(x,y)$, allowing the network to better resolve uncertainty patterns once the primary permeability structures have been learned. Towards the end of training, both weights progressively stabilise, indicating that a balanced regime between the two tasks has been reached. This behaviour demonstrates that GradNorm effectively stabilises the joint learning of permeability magnitude and uncertainty without the need for manual tuning of task-specific loss weights.

\begin{figure}[t!]
\centering
\begin{tikzpicture}
        \node[anchor=north west] at (0,0)
            {
            \begin{subfigure}{.48\textwidth}
            \centering
            \includegraphics[width=\linewidth]{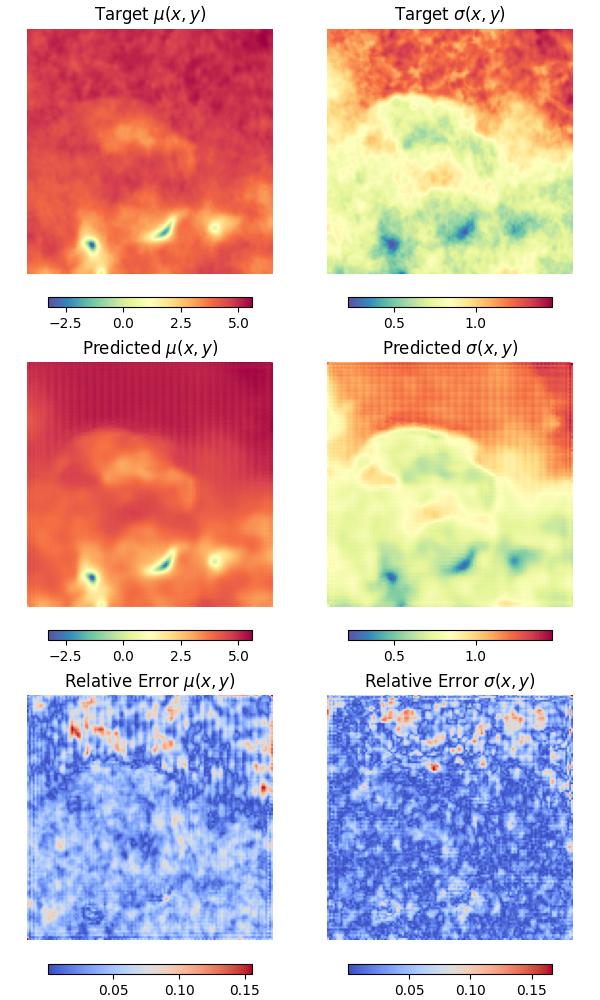}
            \end{subfigure}
            \hspace{1mm}
            \begin{subfigure}{.48\textwidth}
            \centering
            \includegraphics[width=\linewidth]{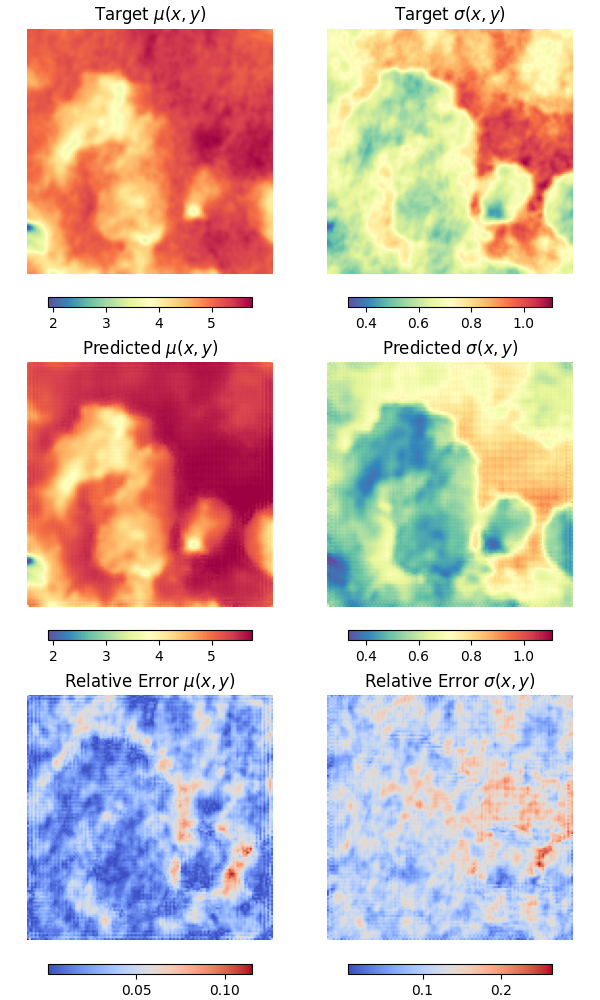}
        \end{subfigure}
            };
        \node at (0.25\linewidth,0.03\linewidth) {Sample A};
        \draw[black, line width=0.5pt] (0.5\linewidth,0) -- (0.5\linewidth,-0.8\linewidth);
        \node at (0.72\linewidth,0.03\linewidth) {Sample B};
    \end{tikzpicture}
\caption{Comparison between target and predicted outputs of the probabilistic permeability mapping for representative samples from the training set. For each sample, the top row shows the target log-normal parameters $\mu(x,y)$ and $\sigma(x,y)$ obtained from the Bayesian inference, the middle row shows the corresponding U-Net predictions, and the bottom row reports the relative error fields.}
\label{fig:UNet_results_train_patches}
\end{figure}

\begin{figure}[t!]
\centering
\begin{tikzpicture}
        \node[anchor=north west] at (0,0)
            {
            \begin{subfigure}{.48\textwidth}
            \centering
            \includegraphics[width=\linewidth]{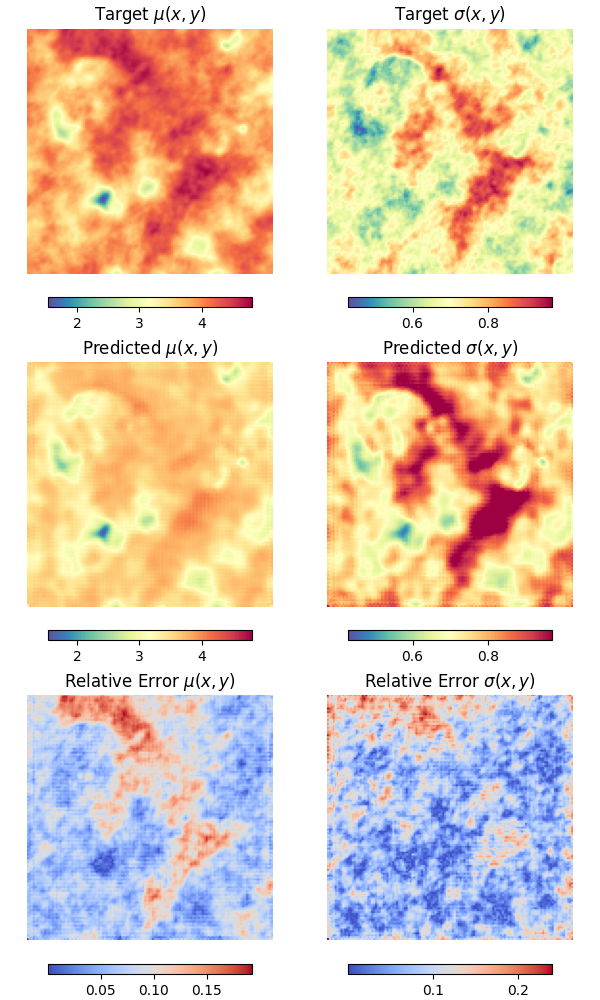}
            \end{subfigure}
            \hspace{1mm}
            \begin{subfigure}{.48\textwidth}
                \centering
                \includegraphics[width=\linewidth]{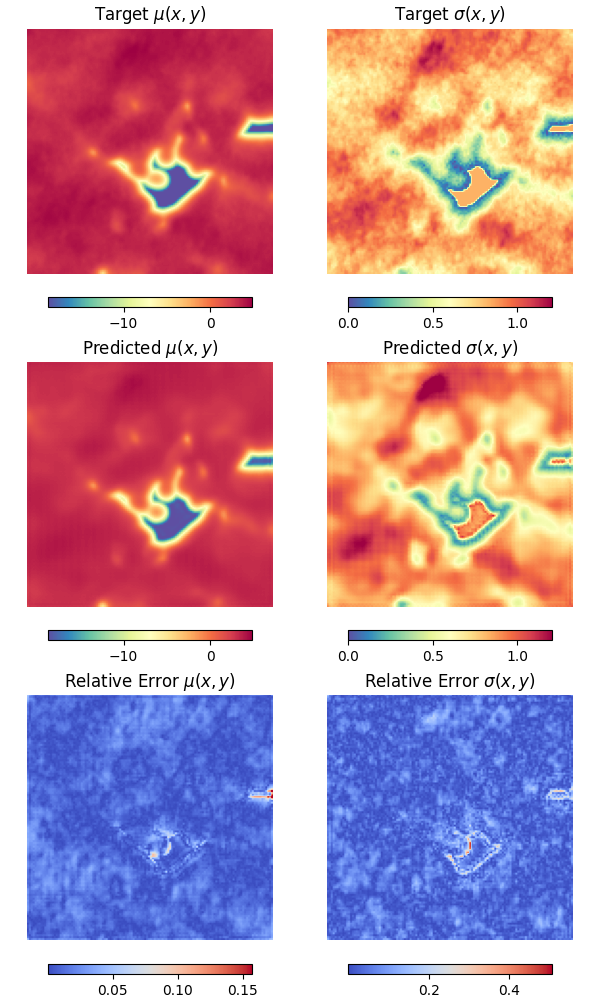}
            \end{subfigure}
            };
        \node at (0.25\linewidth,0.03\linewidth) {Sample A};
        \draw[black, line width=0.5pt] (0.5\linewidth,0) -- (0.5\linewidth,-0.8\linewidth);
        \node at (0.72\linewidth,0.03\linewidth) {Sample B};
    \end{tikzpicture}
\caption{Comparison between target and predicted outputs of the probabilistic permeability mapping for representative samples from the validation set (same format as Figure~\ref{fig:UNet_results_train_patches}). The close agreement between target and predicted fields, together with spatially limited errors, demonstrates the ability of the network to generalise the learned probabilistic mapping beyond the training data.}
\label{fig:UNet_results_val_patches}
\end{figure}

\begin{figure}[ht!]
\centering
\begin{subfigure}{.49\textwidth}
    \caption{}
    \centering
    \includegraphics[width=\linewidth]{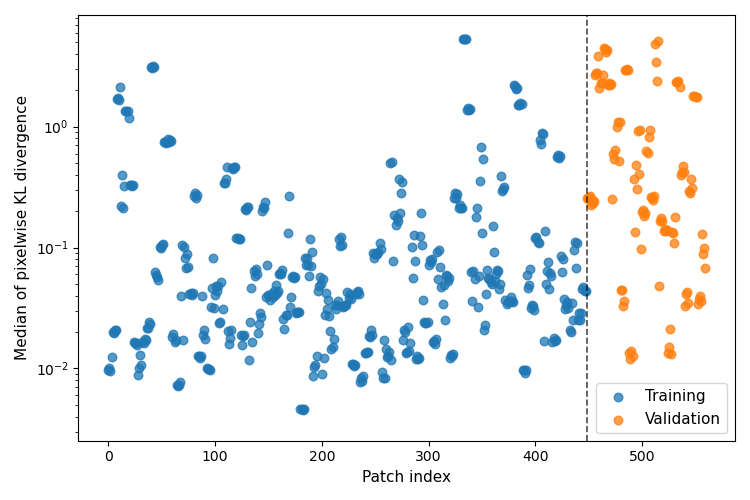}
    \label{fig:dist_comp_median}
\end{subfigure}
\hspace{-1mm}
\begin{subfigure}{.49\textwidth}
    \caption{}
    \centering
    \includegraphics[width=\linewidth]{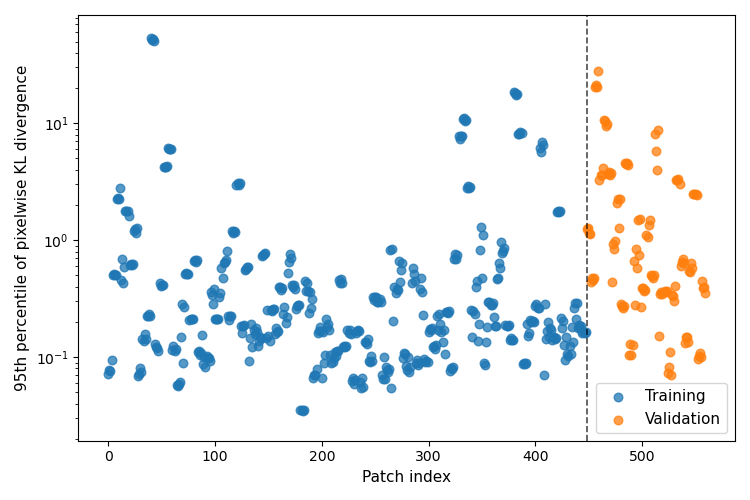}
    \label{fig:dist_comp_percentile}
\end{subfigure}
\caption{Distributional accuracy per patch using symmetric KL divergence. For each patch, the pixelwise KL divergence between the predicted and target log-normal distributions is computed and summarized by (a) the median and (b) the 95th percentile across pixels. Training and validation patches are shown separately, with the dashed vertical line indicating the split between training and validation sets. Most patches exhibit low median divergences (typically $10^{-2}\,$--$\,10^{-1}$), indicating close agreement for the majority of pixels, while the 95th percentile highlights localized discrepancies in more challenging regions.}
\label{fig:dist_comp_val_train}
\end{figure}

The accuracy of the learned mapping is assessed in Figures~\ref{fig:UNet_results_train_patches} and~\ref{fig:UNet_results_val_patches}, which compare the target and predicted log-normal parameters for representative samples from the training and validation sets, respectively. In both cases, the predicted fields reproduce the spatial organisation, amplitude, and connectivity of the target permeability statistics, with relative errors that remain spatially localised and of limited magnitude. We further evaluate distributional accuracy by comparing the predicted and target posteriors through the analysis of the pixelwise symmetric Kullback-Leibler (KL) divergence between the corresponding log-normal distributions. To avoid numerical instabilities in regions where the target posterior collapses, the divergence is evaluated only for pixels satisfying $\sigma(x,y) > 10^{-3}$. For each patch, the resulting divergence field is summarised by its median and 95th percentile across pixels (see Figure~\ref{fig:dist_comp_val_train}). The median divergences are typically in the range $10^{-2}\,$--$\,10^{-1}$, indicating good agreement between predicted and target distributions for most pixels within the patches. The 95th percentile remains mostly below $\mathcal{O}(1)$, with larger values observed in a limited number of patches. These higher divergences correspond to localised, more challenging regions rather than a systematic degradation of performance. Importantly, no systematic increase is observed for the validation samples, indicating that the U-Net generalises the learned probabilistic aperture-permeability relationship beyond the training data and specific aperture patterns.

The evaluation metrics considered here quantify the agreement between the predicted and target posterior distributions at the patch scale. In the present framework, the supervised targets correspond to posterior descriptors obtained from the physics-based Bayesian correction (see Eq.~\ref{eq:am_mu_sigma_mapping}), rather than ground-truth realisations of the permeability field $K(x,y)$. Accordingly, calibration at the patch scale is defined with respect to the inferred posteriors on held-out fracture realisations, and is assessed here through the agreement between the learned posterior distributions and their Bayesian reference. The decisive external validation, however, lies in the physical consistency of the propagated uncertainty at the fracture scale, which is evaluated in Section~\ref{subsec:Darcy_upscaling} through comparison with Stokes effective permeability.

\subsection{Posterior permeability maps on natural fractures}
\label{subsec:Results_Kmap}

The trained U-Net is then applied to the full-scale aperture fields of the natural shear fractures and their corresponding synthetic geometries introduced in Section~\ref{sec:Geological_frac}, yielding spatially resolved probabilistic permeability maps. From the predicted log-normal parameters $\mu(x,y)$ and $\sigma(x,y)$, local permeability distributions are reconstructed and summarised using the descriptors defined in Eq.~\eqref{eq:K_descriptors}, namely the mode, expected value $\mathbb{E}[K(x,y)]$, and lower and upper quantiles quantifying the 95\% confidence interval $\big[Q_{0.025}[K(x,y)],\, Q_{0.975}[K(x,y)]\big]$.

\begin{figure}[ht!]
\centering
\raisebox{0.18\height}{
\begin{subfigure}{.39\textwidth}
    \caption{}
    \centering
    \includegraphics[width=\linewidth]{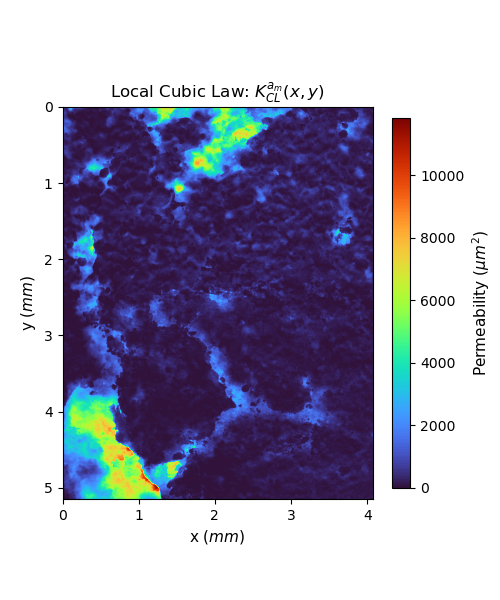}
    \label{fig:LCL_real}
\end{subfigure}
}%
\begin{subfigure}{.59\textwidth}
    \caption{}
    \centering
    \includegraphics[width=\linewidth]{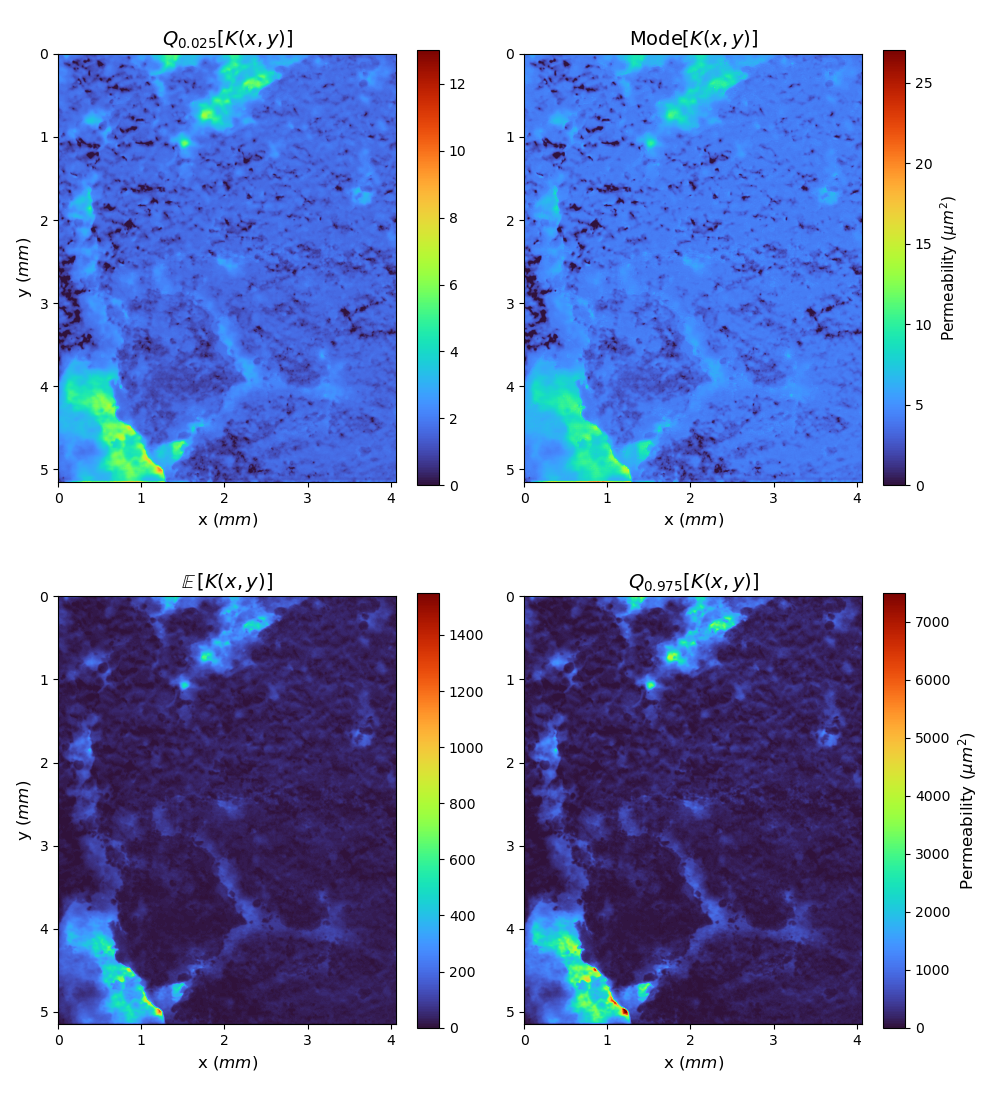}
    \label{fig:Post_K_real}
\end{subfigure}
\caption{Probabilistic permeability maps for the natural shear fracture \#1: (a) Deterministic permeability field obtained from the local Cubic-Law relationship applied to the mechanical aperture, denoted $K_{CL}^{a_m}(x,y)$ following the notation of Perez et al. (2025) \protect\cite{GRL_Perez_2025}. (b) Permeability fields derived from log-normal distribution predicted by the U-Net, including the lower quantile $Q_{0.025}[K(x,y)]$, mode, expected value $\mathbb{E}[K(x,y)]$, and upper quantile $Q_{0.975}[K(x,y)]$, as defined in Eq.~\eqref{eq:K_descriptors}. These maps illustrate how uncertainty in the local aperture-permeability relationship propagates into spatially heterogeneous permeability ranges while preserving the organisation of flow-relevant structures.}
\label{fig:real_results_K}
\end{figure}

Figure~\ref{fig:real_results_K} presents the results for the representative natural shear fracture \#1. For reference, we first show the deterministic permeability field obtained from the local Cubic-Law relationship applied pointwise to the mechanically derived aperture field,
\begin{equation}
    K_{CL}^{\,a_m}(x,y) = \frac{a_m (x,y)^2}{12}.
\label{eq:Perm_map_CL}
\end{equation}
following the notation of Perez et al. (2025) \cite{GRL_Perez_2025}. This field exhibits strong permeability contrasts and pronounced channelisation controlled by the largest local apertures. This local Cubic-Law permeability map is shown here for qualitative comparison and differs from the Darcy-upscaled Cubic-Law permeability $K^{\,a_m}_D$, which is obtained by solving a flow problem on the heterogeneous aperture field \eqref{eq:Perm_map_CL} and is analysed in Section~\ref{subsec:Results_Darcy}. In contrast, the probabilistic permeability maps predicted by the U-Net display a distribution of permeability values at each spatial location, reflecting uncertainty in the local aperture-permeability relationship. While the spatial organisation of high-permeability pathways is preserved across the different probabilistic descriptors, the spread between lower and upper quantiles reveals spatially heterogeneous uncertainty that is not captured by the deterministic Cubic-Law estimate. 

\begin{figure}[t!]
\centering
\raisebox{0.18\height}{
\begin{subfigure}{.39\textwidth}
    \caption{}
    \centering
    \includegraphics[width=\linewidth]{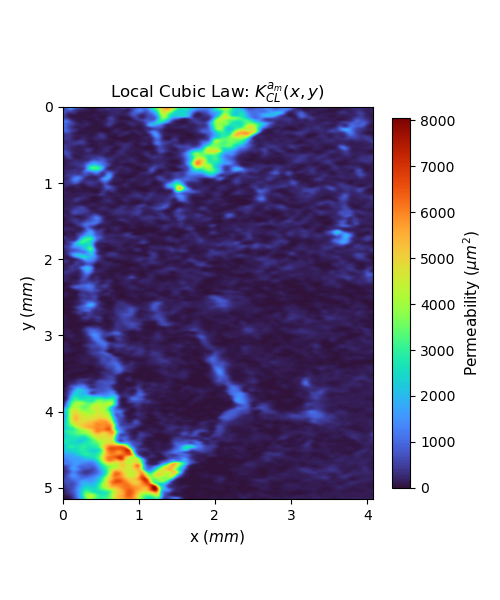}
    \label{fig:LCL_synthetic}
\end{subfigure}
}%
\begin{subfigure}{.59\textwidth}
    \caption{}
    \centering
    \includegraphics[width=\linewidth]{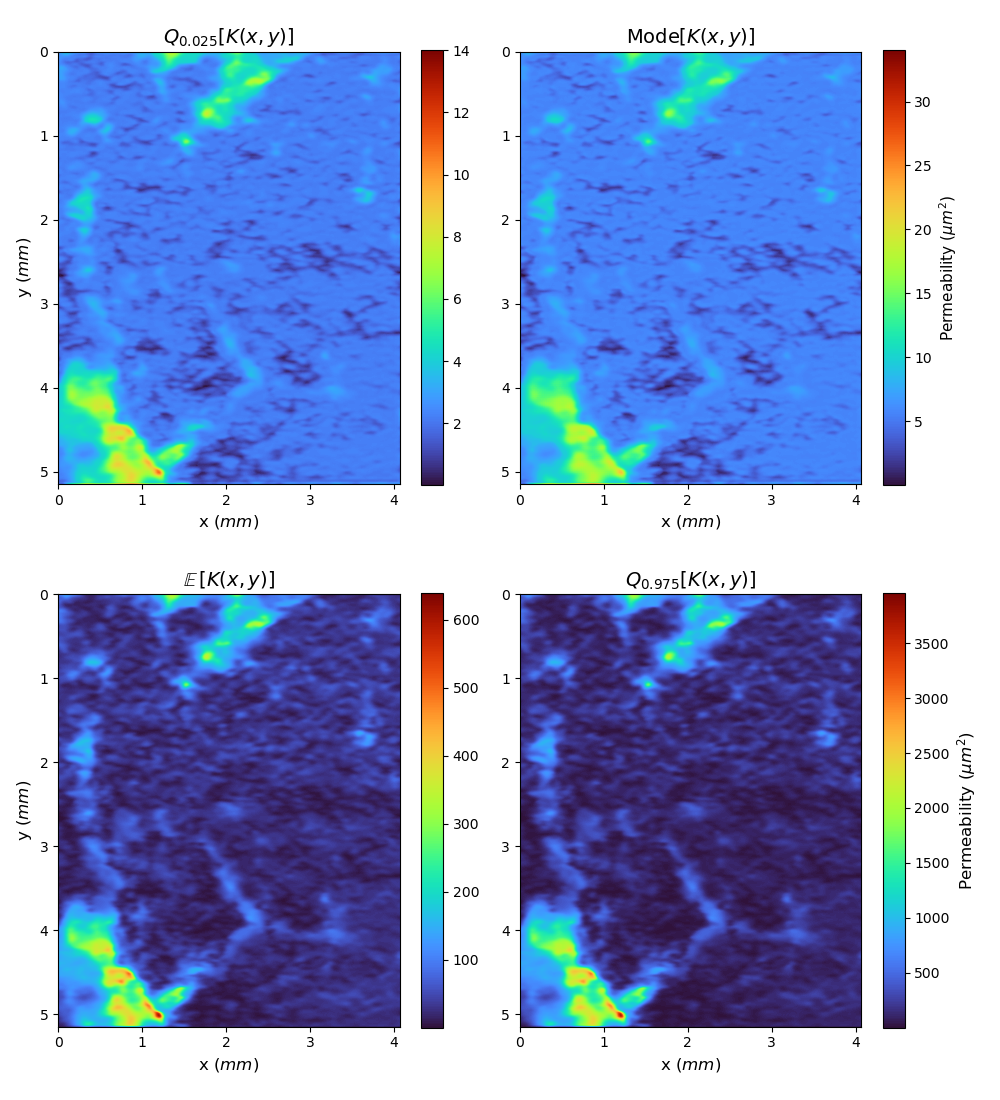}
    \label{fig:Post_K_synthetic}
\end{subfigure}
\caption{Probabilistic permeability maps for the synthetic fracture \#1. (a) Deterministic permeability field obtained from the local Cubic-Law, $K_{CL}^{a_m}(x,y)$. (b) Probabilistic permeability descriptors predicted by the U-Net, including the lower quantile, mode, expected value, and upper quantile. Compared to the natural fracture (see Figure~\ref{fig:real_results_K}), the synthetic fracture exhibits smoother spatial patterns and a narrower spread between permeability quantiles.}
\label{fig:synthetic_results_K}
\end{figure}

\begin{figure}[ht!]
    \centering
    \includegraphics[width=\linewidth]{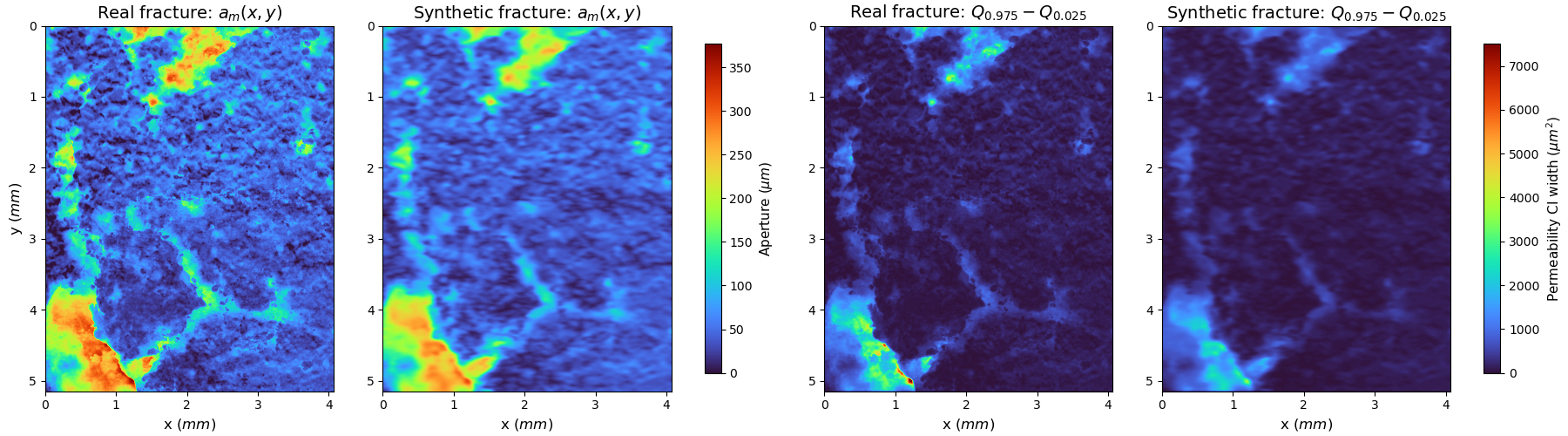}
    \caption{Thickness-weighted mechanical aperture fields $a_m(x,y)$ and corresponding local permeability uncertainty widths, quantified as the difference between the upper and lower quantiles for the natural fracture \#1 and its synthetic counterparts. The natural fracture exhibits strongly localised uncertainty patterns associated with abrupt aperture transitions, whereas the synthetic fracture displays a smoother aperture field and a more homogeneous, reduced uncertainty range.}
\label{fig:Ap_vs_UQ_width}
\end{figure}

The width of the 95\% credible interval, highlighted in Figure~\ref{fig:Ap_vs_UQ_width}, reveals a spatially heterogeneous uncertainty field. Elevated uncertainty is localised in regions where the aperture field exhibits abrupt transitions and channel intersections. This behaviour reflects the increasing mismatch between simplified aperture–permeability relationships and the true hydraulic response in geometrically complex regions. In particular, roughness and channelling locally violate the assumptions underlying Cubic-Law models, leading to increased modelling uncertainty, which is associated with higher posterior variance in the physics-based Bayesian correction step (see Section~\ref{subsec:Bayesian_inference}). In contrast, regions characterised by smoother spatial variations in aperture, i.e. lower local gradients or longer correlation lengths, exhibit reduced uncertainty, as the local hydraulic response is better approximated by simplified constitutive relationships. Compared to the natural fracture, the synthetic fracture displays a narrower and smoother uncertainty distribution (see Figure~\ref{fig:synthetic_results_K} and~\ref{fig:Ap_vs_UQ_width}), suggesting that the reduction in geometric complexity is associated with reduced variability in the permeability uncertainty. 

Equivalent probabilistic permeability maps and uncertainty fields for the natural shear fracture \#2 are provided in~\ref{app:NF2} (Figure~\ref{fig:real_results_K_NF2} and~\ref{fig:Ap_vs_UQ_width_NF2}). The spatial organisation of high-permeability pathways and uncertainty patterns remains consistent with those observed for fracture \#1. In contrast to fracture \#1, however, the differences between the real and synthetic representations are less pronounced for fracture \#2, reflecting its reduced branching complexity. This interpretation is supported by the geometric analysis in Section~\ref{subsec:fracture_comparison}, which shows similar correlation lengths and a less complex aperture topology for fracture \#2.

\begin{figure}[ht!]
    \centering
    \includegraphics[width=\linewidth]{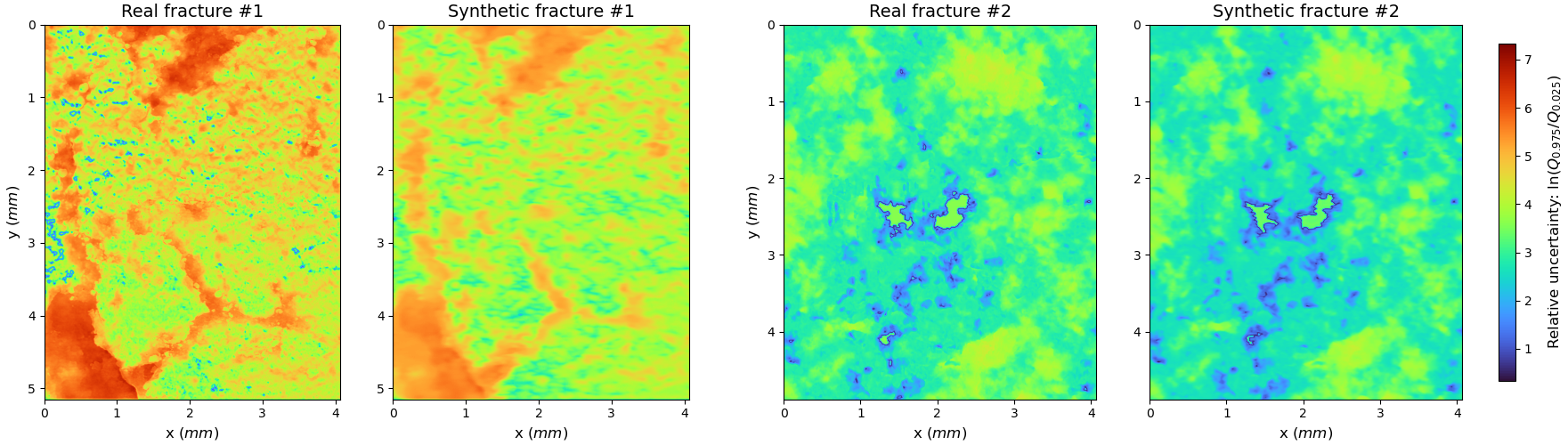}
\caption{
Relative uncertainty maps of the predicted local permeability distributions for the natural shear fractures \#1 and \#2 and their synthetic counterparts. The local relative uncertainty is represented by the logarithm of the quantile ratio $\ln(Q_{0.975}/Q_{0.025})$, which is directly proportional to the predicted log-normal standard deviation $\sigma(x,y)$ and therefore provides a scale-independent comparison of uncertainty across fractures.
}\label{fig:UQ_width_comp}
\end{figure}

However, because the two natural fractures exhibit markedly different permeability scales (see Table~\ref{table2}), direct comparison of absolute uncertainty widths is not meaningful. To enable a consistent comparison, we consider a scale-independent metric based on the logarithm of the quantile ratio, $\ln(Q_{0.975}/Q_{0.025})$, which is directly proportional to the predicted log-normal standard deviation $\sigma(x,y)$ (see Figure~\ref{fig:UQ_width_comp}). This metric provides relative uncertainty maps, showing that fracture \#2 exhibits systematically lower uncertainty levels, reflecting its larger mean aperture and reduced relative roughness (see Table~\ref{table1}), which lead to a more homogeneous hydraulic response and reduced variability in the inferred permeability field. While this indicates a reduction in local uncertainty, it does not imply a proportional reduction in uncertainty at the fracture scale, as effective permeability is also controlled by the spatial organisation and connectivity of flow pathways. As a result, even moderate local uncertainties can lead to significant variability in the upscaled hydraulic response.

Taken together, these results demonstrate that the proposed framework produces permeability maps that are spatially structured, uncertainty-aware and scalable to realistic geometries. The probabilistic descriptors predicted by the U-Net provide a physically interpretable alternative to deterministic aperture-permeability mappings, capturing how geometric complexity translates into local uncertainty while preserving the connectivity of flow-relevant features. Translating spatially heterogeneous and uncertain permeability fields into large-scale hydraulic responses finally requires a flow-based upscaling step that accounts for such connectivity and non-linear effects in the uncertainty propagation.

\subsection{Upscaled uncertainties on the hydraulic response}
\label{subsec:Results_Darcy}

The probabilistic permeability fields obtained in Section~\ref{subsec:Results_Kmap} provide a spatially resolved description of local hydraulic properties together with their uncertainty. To evaluate how these local uncertainties propagate to the fracture scale, we perform flow-based upscaling using steady-state Darcy simulations, following the procedure described in Section~\ref{subsec:Darcy_upscaling}. This approach enables a direct quantification of how uncertainty in the local aperture-permeability relationship translates into uncertainty in the effective fracture permeability $K_{\mathrm{eff}}$. 

\begin{figure}[ht!]
    \centering
    \includegraphics[width=0.8\linewidth]{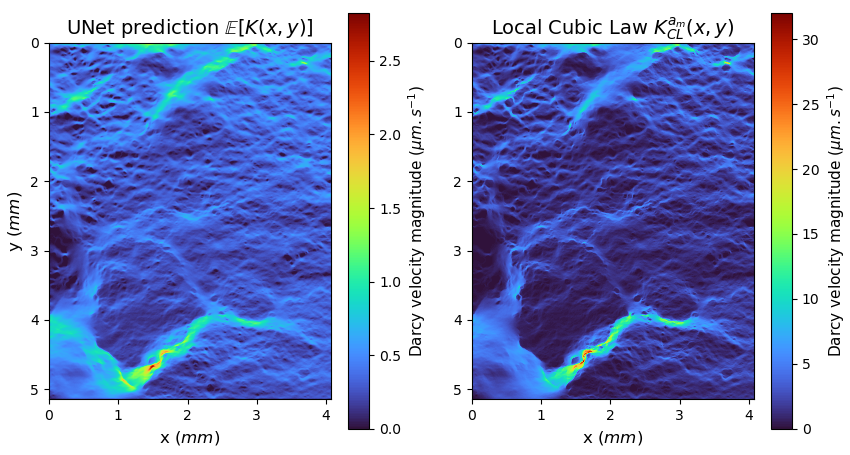}
    \caption{
    Comparison of Darcy velocity magnitude fields for the natural shear fracture \#1 obtained using two permeability models:
    (a) Velocity field computed from the probabilistic permeability predicted by the U-Net, using the expected value $\mathbb{E}[K(x,y)]$ of the local log-normal distribution.
    (b) Velocity field computed from a deterministic permeability derived from the local Cubic-Law applied to the mechanical aperture, $K^{\,a_m}_{CL}(x,y)$. Both simulations are performed under identical boundary conditions. While the main flow pathways are similarly located, the Cubic-Law model produces stronger velocity contrasts and higher peak velocities, whereas the U-Net-based permeability yields smoother velocity distributions that reflect uncertainty-aware redistribution of flow through the fracture.}    
\label{fig:Darcy_velocities_mag}
\end{figure}

Figure~\ref{fig:Darcy_velocities_mag} and~\ref{fig:Darcy_velocities_mag_NF2} (in~\ref{app:NF2}) illustrate the impact of permeability modelling assumptions on the spatial distribution of flow within the natural shear fractures \#1 and \#2, respectively. Darcy velocity fields computed using the U-Net-predicted expected permeability field, $\mathbb{E}[K(x,y)]$, reproduce the main preferential flow pathways while exhibiting smoother velocity gradients and a more diffuse distribution of flux. In contrast, velocities obtained from the deterministic Cubic-Law permeability $K_{CL}^{a_m}(x,y)$ are more strongly localised within wide-aperture channels and show sharper contrasts between conductive and poorly connected regions. These differences arise because deterministic aperture-permeability mappings emphasise local aperture extremes, whereas the probabilistic framework redistributes flow across competing pathways. More importantly, these differences directly influence the propagation of uncertainty at the fracture scale, as flow redistribution can amplify or attenuate the impact of local permeability variability based on the connectivity and organisation of flow pathways.

Beyond these spatial patterns, the key implication thus lies in how these differences affect fracture-scale hydraulic responses. The probabilistic envelopes for the fracture-scale hydraulic responses are defined by the 95\% confidence intervals $\big[K_{Q_{0.025}},\, K_{Q_{0.975}}\big]$, with log-normal distributions parametrising the upscaled uncertainties on the effective permeabilities. Table~\ref{table2} summarises the resulting upscaled permeabilities for the natural shear fractures and their synthetic counterparts, together with deterministic reference values. The latter include the Cubic-Law approximations, $K_{CL}$ and $K^{\,a_m}_D$, and reference Stokes permeabilities $K_{NS}$. It is important to note that the deterministic Stokes permeability is conditional on a specific realisation of the fracture geometry obtained from segmentation. In practice, uncertainties in image processing and segmentation can lead to multiple plausible void-space geometries, and thus to different aperture fields, each associated with a distinct Stokes solution. A rigorous quantification of such variability would require an ensemble of Stokes simulations over perturbed geometries, which is computationally prohibitive. The probabilistic framework proposed here provides an efficient alternative by capturing both data uncertainty in the aperture field and model discrepancies associated with reduced-order flow representations, without requiring repeated high-fidelity simulations. For the natural fractures, deterministic Cubic-Law approaches significantly overestimate transmissivity relative to the Stokes reference, consistent with the findings of Perez et al. (2025) \cite{GRL_Perez_2025}. In contrast, the probabilistic upscaled permeability distribution predicted by the U-Net encompasses the Stokes-consistent values within their uncertainty ranges, with the mode being closer to $K_{NS}$ than any Cubic-Law estimates. The synthetic fractures exhibit a similar behaviour but with narrower uncertainty ranges. While their deterministic Cubic-Law permeabilities are also higher than the Stokes references, the separation between probabilistic quantiles is reduced compared to the natural fractures. This reflects the lower gradients, i.e. smoother spatial variations in the aperture field, and longer correlation lengths of the synthetic fractures, which limit flow redistribution and channel competition. Importantly, the probabilistic framework does not enforce reduced uncertainty a priori, but the narrower uncertainty ranges rather emerge naturally from the simpler geometry of the aperture field, inherited from the underlying void structure..

\begin{figure}[pb!]
\centering
\begin{subfigure}{.49\textwidth}
    \caption{}
    \centering
    \includegraphics[width=\linewidth]{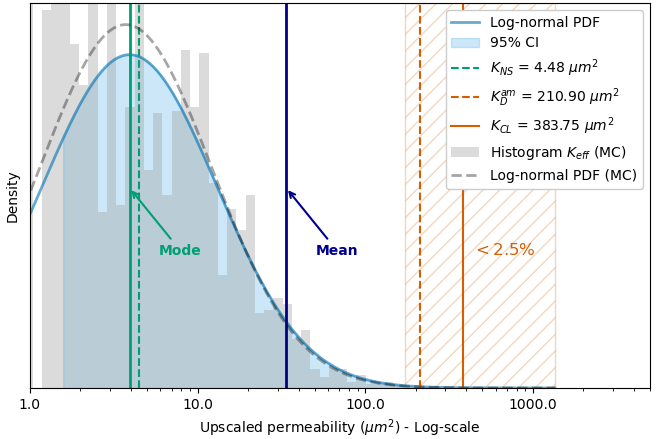}
    \label{fig:Up_K_real}
\end{subfigure}
\begin{subfigure}{.49\textwidth}
    \caption{}
    \centering
    \includegraphics[width=\linewidth]{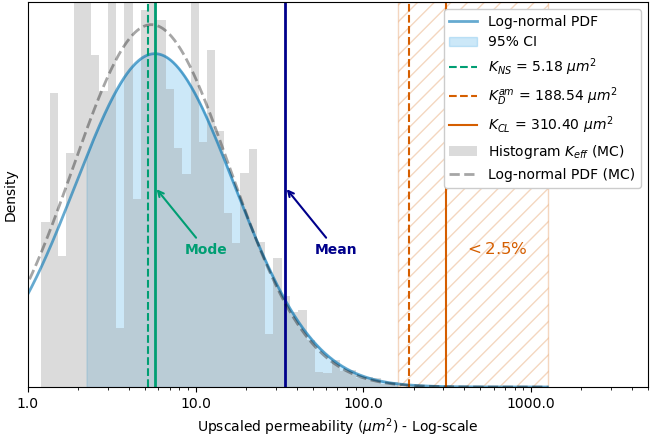}
    \label{fig:Up_K_Synthetic}
\end{subfigure}
\caption{Probability distributions of the upscaled effective permeability $K_{\mathrm{eff}}$ for (a) the natural shear fracture \#1 and (b) its synthetic counterpart. Analytical distributions $\mathcal{P}$ (in blue) are reconstructed from the Darcy-upscaled permeability moments $K_\mathrm{mode}$ and $K_\mathrm{mean}$ predicted by the U-Net, and compared with the Monte Carlo distributions $\mathcal{P}_{\mathrm{MC}}$ (in grey) for 400 permeability realisations (see Eq.~\eqref{eq:perm_real_MC}). The shaded blue area represents the 95\% confidence interval of the analytical distribution, with the solid green and blue lines indicating the mode and mean, respectively. The orange hatched region marks the upper 2.5\% tail of the distribution, highlighting rare and extreme high-value realisations. Vertical markers represent deterministic reference permeabilities obtained from Cubic-Law assumptions (in orange) and Stokes-consistent models (in green).}
\label{fig:Up_K_results}
\end{figure}

\begin{figure}[pb!]
\centering
\begin{subfigure}{.49\textwidth}
    \caption{}
    \centering
    \includegraphics[width=\linewidth]{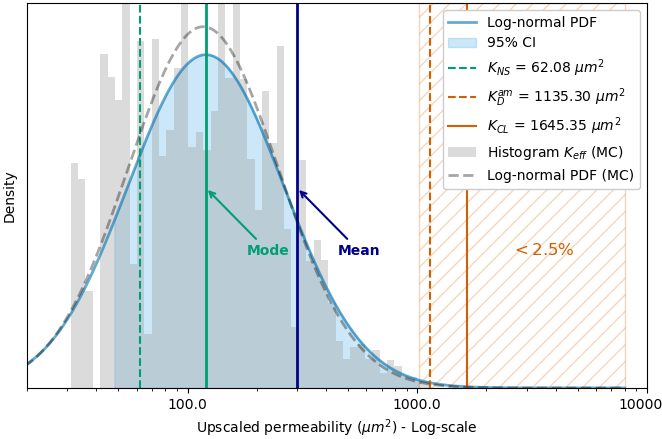}
    \label{fig:Up_K_real_NF2}
\end{subfigure}
\begin{subfigure}{.49\textwidth}
    \caption{}
    \centering
    \includegraphics[width=\linewidth]{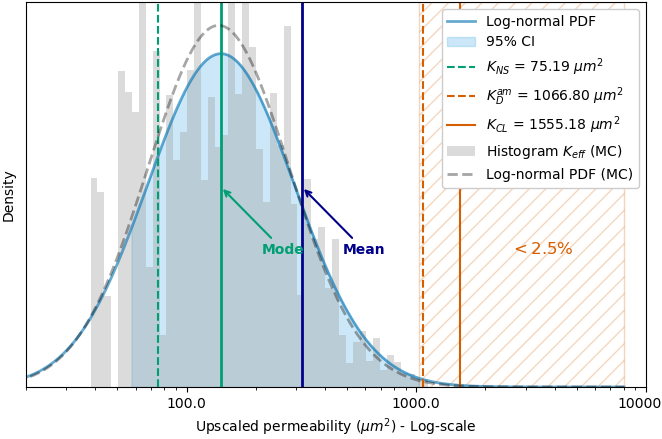}
    \label{fig:Up_K_Synthetic_NF2}
\end{subfigure}
\caption{Probability distributions of the upscaled effective permeability $K_{\mathrm{eff}}$ for (a) the natural shear fracture \#2 and (b) its synthetic counterpart. Analytical distributions reconstructed from the predicted log-normal parameters are compared with Monte Carlo estimates. The shaded blue region represents the 95\% confidence interval of the analytical distribution, while vertical lines denote the mean, mode, and reference permeability values.}
\label{fig:Up_K_results_NF2}
\end{figure}

Figure~\ref{fig:Up_K_results} summarises these results  for fracture \#1 by showing the reconstructed probability distributions of the effective permeability $K_\mathrm{eff}$. The close agreement between the analytical distribution $\mathcal{P}$ derived from the descriptor fields (Section~\ref{subsec:Darcy_upscaling}) and the Monte Carlo propagation $\mathcal{P_{\mathrm{MC}}}$ based on the fully correlated latent-variable model confirms that the descriptor-based representation provides a reliable approximation of uncertainty propagation at the fracture scale. This agreement is further quantified through the Kullback–Leibler divergence $D_{\mathrm{KL}}\left(\mathcal{P_{\mathrm{MC}}} || \mathcal{P}\right) = \mathcal{O}(10^{-3})$ which remains very small for the natural fractures and their synthetic counterparts, indicating that both distributions carry nearly identical information. The natural shear fracture exhibits a broader and more skewed distribution, reflecting strong spatial heterogeneity and channel competition (see Figure~\ref{fig:Up_K_real}). In contrast, the synthetic fracture yields a more concentrated distribution with reduced variance (see Figure~\ref{fig:Up_K_Synthetic}). In both cases, the Stokes-consistent permeability falls within the probabilistic envelope, whereas deterministic Cubic-Law estimates lie outside or near the upper tail of the distribution. Results on the effective permeability distributions for the second natural fracture are reported in Figure~\ref{fig:Up_K_results_NF2}, and exhibit similar trends as observed for fracture \#1, with Stokes-consistent permeabilities falling within the predicted probabilistic envelopes, while Cubic-Law estimates largely overestimate transmissivity.

These results demonstrate that explicitly accounting for uncertainty in the local aperture-permeability relationship leads to fracture-scale permeability estimates that are both physically consistent and quantitatively compatible with high-fidelity Stokes simulations. Rather than relying on a single deterministic aperture proxy, our probabilistic framework captures how geometric complexity, channel competition, and flow redistribution jointly control fracture transmissivity. The contrast between natural shear fractures and their simplified synthetic counterparts further highlights how fracture geometry governs not only mean hydraulic behaviour but also the spread of admissible macroscopic responses. These findings establish a robust basis for propagating fracture-scale uncertainty to larger-scale fracture networks.

\begin{table}[t!]
\centering
\begin{tabular}{cccccc}
\hline \multirow{2}{*}{Upscaled Permeabilities} & \multirow{2}{*}{Units} &\multicolumn{2}{c}{Fracture \#1} & \multicolumn{2}{c}{Fracture \#2} \\
 & & Real & Synthetic & Real & Synthetic\\
\hline 
$K_{CL}$ & \multirow{3}{*}{$\mu m^2$} & 383.8 & 310.4 & 1645.4 & 1555.2 \\
$K^{\, a_m}_{D}$ & & 210.9 & 188.5 & 1135.3 & 1066.8 \\
$K_{NS}$ & & 4.48 & 5.18 & 62.08 & 75.19  \\
\hline
$K_{Q_{0.025}}$ & \multirow{3}{*}{$\mu m^2$} & 1.58 & 2.22 & 47.83 & 57.32 \\
$K_{\mathrm{mode}}$ & & 3.93 & 5.72 & 120.1 & 141.1  \\
$K_{\mathrm{mean}}$ & & 33.78 & 34.14 & 299.7 & 317.9 \\ 
$K_{Q_{0.975}}$ & & 172.5 & 159.8  & 1020.8 & 1025.9 \\
\hline 
\end{tabular}
\caption{Upscaled effective permeabilities $K_{\mathrm{eff}}$ obtained from Darcy flow simulations on deterministic and probabilistic permeability fields for the natural shear fractures and their synthetic counterparts. Reported deterministic values include classical Cubic-Law estimates $K_{CL}$, Darcy-upscaled local Cubic-Law permeabilities based on the mechanical aperture field $K^{\,a_m}_D$, Stokes-consistent permeabilities $K_{NS}$. The probabilistic effective permeabilities corresponds to the lower quantile $K_{Q_{0.025}}$, mode $K_{\mathrm{mode}}$, mean $K_{\mathrm{mean}}$, and upper quantile $K_{Q_{0.975}}$ as defined in Eq.~\eqref{eq:upscaled_perm_ranges} in Section~\ref{subsec:Darcy_upscaling}. All values are reported in the main flow direction.}
\label{table2}
\end{table}

\section{Discussion}
\label{sec:discussion}
We demonstrated that the proposed probabilistic upscaling framework captures both the spatial organisation of flow and the associated uncertainty for complex natural fractures. Here, we discuss these findings in terms of (i) the ability of the learned permeability mapping to generalise from local training patches to full-scale natural fracture geometries  (Section~\ref{subsec:generalisation}); (ii) the computational efficiency gained by replacing repeated Stokes-based simulations with a data-driven surrogate for fast uncertainty quantification analysis (Section~\ref{subsec:comp_eff}); and (iii) the potential extension of the approach to fracture networks and larger-scale models (Section~\ref{susec:frac_net_extension}). 

\subsection{Generalisation}
\label{subsec:generalisation}

The U-Net model (Figure~\ref{fig:unet_architecture}) is trained exclusively on local aperture patches extracted from $\mu$CT volumes, while its ultimate application targets full-scale natural fracture geometries characterised by long-range correlations, anisotropy, and multi-scale channelisation. An important question is therefore whether the local data adequately represent the spatial statistics governing fracture-scale hydraulic behaviour, and whether the model can generalise beyond the training regime.

To address this, we compare the spatial correlation lengths of the natural fractures $\xi_{frac}$ with those present in the training dataset $\xi_{loc}$. We follow the same procedure as in Section~\ref{subsec:fracture_comparison}: directional variograms of $a_m(x,y)$ are computed and fitted with an exponential model, and the correlation lengths are defined as the lags at which the fitted variograms reach 95\% of its sill. Correlation lengths results for the natural shear fractures are reported in Table~\ref{table1} and range from $1.49$ to $1.55\, \mathrm{mm}$. By comparison, the characteristic length of the training patches is $L_{loc}\approx0.35~\mathrm{mm}$. When applying the same variogram-based analysis patchwise, approximately 55\% of the extracted samples exhibit local correlation lengths $\xi_{loc}<L_{loc}$, indicating that a substantial fraction of the small-scale heterogeneity is fully resolved within individual patches. These samples capture contact zones, sharp aperture contrasts, and local channelisation associated with short-range roughness that govern preferential flow pathways at pore to sub-millimetre scales. The remaining $45\%$ of patches display correlation lengths comparable to or exceeding the patch size, reflecting smoother aperture variations dominated by larger-scale trends. Although such samples do not fully resolve long-range structures present in the natural fractures, they inform about extended conductive zones and gradual aperture gradients. This combination of highly heterogeneous and trend-dominated patches exposes the network to a broad spectrum of spatial configurations, limiting overfitting to specific local patterns and promoting scale-robust feature learning.

\begin{figure}[pb!]
    \centering
    \includegraphics[width=0.8\linewidth]{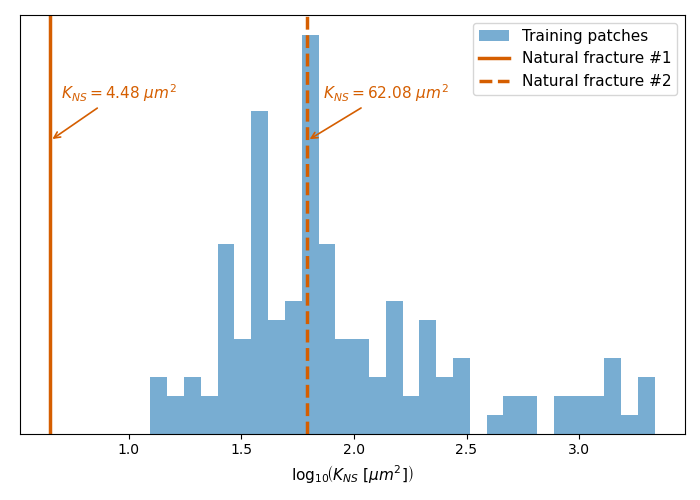}
    \caption{Distribution of Stokes permeabilities $K_{NS}$ obtained on the local aperture patches used for U-Net training (blue histogram shown in $\log_{10}(K_{NS}[\mu m^2])$ space). The vertical lines indicate the permeability values of the natural shear fractures \#1 and \#2, with $K_{NS}=4.48\,\mu m^2$ and $K_{NS}=62.08\,\mu m^2$, respectively. The upscaled permeability of fracture \#1 lies outside the range spanned by the training patches, indicating efficient generalisation beyond the training distribution.}
\label{fig:perm_generalisation}
\end{figure}

The ability of the trained model to generalise is further supported by comparing the permeability distribution obtained on the training set with the Stokes-consistent permeability of the natural fractures. Figure~\ref{fig:perm_generalisation} shows that the upscaled permeability $K_{NS}$ of fracture \#1 lies outside the range spanned by the training patches, indicating the model ability to  extrapolate beyond the permeability range encountered during the training. The fact that the model remains consistent with the Stokes reference at the fracture scale (Section~\ref{sec:Results}) therefore supports its ability to generalise beyond the local training regime and to transfer the learned features to previously unseen permeability levels and fracture geometries.

Anisotropy in fracture hydraulics can also be recovered directly within the proposed framework because the training dataset is constructed under a prescribed flow direction (see Section~\ref{subsec:training_dataset}). As a result, directional permeability for a fracture can be obtained by rotating the aperture map and applying the same trained network to the rotated field. This effectively exchanges the flow-parallel and transverse directions without altering the underlying aperture statistics, allowing the anisotropic response to be characterised. Finally, this avoids solving the full 3D Stokes problem for multiple flow directions, contributing to both the generalisation and computational efficiency of the framework.

Overall, these results indicate that generalisation arises from learning statistically representative relationships between aperture structure and permeability response, rather than from reproducing particular fracture-scale patterns. This ability to generalise from local scale training to fracture-scale predictions underpins the robustness of the probabilistic permeability predictions presented in Section~\ref{sec:Results}.

\subsection{Computational efficiency}
\label{subsec:comp_eff}

Computational efficiency constitutes a major practical advantage of the proposed uncertainty propagation framework. Once trained, the U-Net predicts the full posterior permeability distribution over large natural fracture domains within a few seconds. The subsequent Darcy-scale simulations required to compute fracture-scale confidence intervals are also completed in a few minutes. In contrast, fully resolved 3D Stokes simulations on the same $\mu$CT-derived geometries typically require several hours for a single deterministic permeability estimate. As a result, mechanically inferred aperture maps can be converted into probabilistic hydraulic responses, including uncertainty bounds on $K_{\mathrm{eff}}$, within minutes.

This efficiency gain arises from the use of an effective quasi-2D hydraulic representation combined with data-driven permeability inference, rather than from a direct approximation of 3D flow. High-fidelity Stokes simulations on realistic fracture geometries are prohibitively expensive for uncertainty quantification, particularly when repeated evaluations are needed to explore aperture variability, measurement uncertainty, or model misspecification. Such computational costs limit large-scale Monte Carlo analyses or systematic sensitivity studies. In contrast, the present framework replaces repeated 3D flow simulations with a learned mapping from mechanically derived aperture fields to spatially resolved in-plane permeability distributions, which remain consistent with fracture-scale hydraulic response. Uncertainty propagation is then performed using inexpensive Darcy-scale simulations, while avoiding to rely on empirical aperture-permeability laws.

By decoupling local permeability inference from expensive flow solvers, the proposed approach enables efficient ensemble simulations and uncertainty analyses on realistic fracture geometries. This computational scalability is essential for extending uncertainty-aware hydraulic modelling to large fracture datasets and, ultimately, to fracture network systems. This framework therefore provides a practical pathway for integrating high-resolution imaging, probabilistic inference, and large-scale flow modelling within a unified and tractable workflow.

\subsection{Extension to fracture networks}
\label{susec:frac_net_extension}

Natural subsurface flow systems are commonly controlled by interconnected fracture networks rather than isolated fractures, such as in geological faults. A key objective of uncertainty-aware hydraulic modelling is therefore to enable robust predictions at the network fracture scale in reservoirs and seals. The present framework is well suited for such extensions, as it provides probabilistic permeability descriptors together with physically consistent uncertainty bounds at low computational cost.

A first and natural direction is to use the upscaled probabilistic permeability on $K_{\mathrm{eff}}$ as input to discrete fracture network (DFN) representations \cite{hyman_dfnworks_2015}. In a DFN model, such probabilistic properties can be assigned directly to network components (\emph{e.g.}, as transmissivities or conductances), and uncertainty can be propagated to network responses either through ensemble sampling or by evaluating a small number of representative parameter sets \cite{Hyman_2016, Berrone_2018}. From a computational perspective, the efficiency of the U-Net inference and Darcy-scale upscaling makes large ensemble simulations on fracture networks feasible. Thousands of stochastic realisations can be generated by sampling from the predicted permeability distributions, providing a practical pathway for uncertainty quantification of network-scale transmissivity, breakthrough times, connectivity and transport metrics \cite{MAKEDONSKA_2016, Philippe_Davy_PNAS_2024}. 

A second, complementary approach is to move beyond the classical DFN assumption of strictly planar, 2D fracture elements, and instead employ a reduced 3D Darcy representation for regions of the network where the void space forms a fracture zone rather than a single surface (\emph{e.g.}, branching segments or intersections). Building on the thickness-weighted geometric reduction introduced here, complex local 3D geometries can be partitioned into subdomains and mapped to a set of 2D observables via thickness-weighted averaging. In this hybrid view, the reduced 3D acts as an intermediate-scale closure that captures connectivity effects within complex network elements while maintaining an uncertainty-aware description that can be coupled back into network-scale flow and transport models. Furthermore, this approach can preserve the computational efficiency of the current uncertainty propagation strategy based on representative descriptor fields (mean, mode and quantiles), hence avoiding large ensembles of DFN simulations.

Finally, fracture network applications introduce additional variability in local boundary conditions and confinement, particularly near intersections where openings may be partially confined or unconfined. This motivates augmenting the aperture-based input with additional geometric indicators extracted from the 3D segmentation (\emph{e.g.}, open-thickness fraction or distance-to-wall measures) to ensure that predicted probabilistic permeability remains consistent across confinement regimes. In the present work, only well-confined fracture configurations, consistent with the $\mu$CT-derived shear fractures, have been analysed.


\section{Conclusion}

This work presents a scalable probabilistic upscaling workflow for flow through geological fractured media that bridges image-based fracture geometry and uncertainty-aware hydraulic predictions across scales. The central contribution is a hybrid strategy that preserves physical consistency while enabling efficient uncertainty propagation. Specifically, we combine: (i) a Bayesian correction step that accounts for misspecification in aperture-permeability relations derived from uncertain mechanical aperture measurements \cite{GRL_Perez_2025}, (ii) a patch-based deep learning surrogate that maps local aperture structure to spatially distributed permeability statistics, and (iii) Darcy-scale flow upscaling to translate local uncertainty into effective permeability bounds. A key outcome is that uncertainty is explicitly propagated from aperture measurements and constitutive assumptions to macroscopic flow responses, rather than being reduced to a single deterministic permeability estimate. The Bayesian component represents epistemic uncertainty associated with imperfect constitutive assumptions, while the Residual U-Net quantifies the effects of heterogeneity and spatial correlation by operating on local aperture-permeability patterns. Applied to natural shear fractures and to simplified synthetic geometries, the framework reproduces the characteristic impact of channelisation, connectivity, and multivalued void space on macroscopic transmissivity while quantifying the resulting range in admissible hydraulic responses.

Scalability arises because the computationally intensive physics is confined to the preprocessing (supported by a limited set of reference simulations), whereas subsequent training and inference enable efficient ensemble generation of macroscopic predictions suitable for uncertainty propagation and sensitivity analysis. The computational implications are important for practical fracture characterization, since high-fidelity Stokes simulations are prohibitive for large ensembles of fractures. In contrast, once trained, the surrogate produces spatially resolved probabilistic descriptors at negligible marginal cost, and the Darcy upscaling step provides a consistent macroscopic response that preserves the influence of preferential pathways. This makes it feasible to (i) quantify uncertainty for many fractures in a dataset, (ii) compare alternative geometric representations and preprocessing choices, and (iii) propagate uncertainty beyond individual fractures without repeated high-resolution flow computations. Furthermore, the patch-based learning strategy adopted here facilitates transferability to heterogeneous fracture populations. The network can be applied to fractures of varying size, orientation, and geological origin. This supports probabilistic upscaling from core-scale imaging to field-scale fracture networks. For instance, the present outputs can be interpreted as probabilistic transmissivity inputs for discrete fracture network models, with uncertainty bounds that can be used to quantify network-scale variability in equivalent permeability and flow partitioning.

Future developments may integrate the present methodology with geomechanical models, time-dependent aperture evolution, and multiphase or reactive transport processes, as well as uncertainty-aware representations of fault permeability \cite{Viswanathan_2022, Salgado_2022}. Coupling with stress-dependent closure and shear dilation would enable time-varying transmissivity predictions under evolving boundary conditions, while multiphase extensions would allow uncertainty-aware assessment of relative permeability, phase interference, and capillary effects in natural fractures \cite{Phillips2024, Adenike_2015}. Such advances would support uncertainty quantification under coupled hydromechanical and reactive conditions, ultimately improving prediction of fractured geological media behaviour in geothermal reservoirs, geological \ce{CO2} storage, subsurface energy storage or radioactive waste disposal. 

\appendix
\renewcommand{\thesection}{Appendix \Alph{section}:}
\renewcommand{\thesubsection}{\thesection.\arabic{subsection}}

\section{GradNorm weighting for multi-output training}
\label{app:gradnorm}

The Residual U-Net is trained to regress the two log-normal permeability parameters $\mu(x,y)$ and $\sigma(x,y)$, using the multi-objective loss defined in Eq.~\ref{eq:loss_function}. An adaptive task weighting strategy, based on the GradNorm criterion, is used to balance the training \cite{Chen2017GradNorm}. Here, we provide additional details specific to this weighting strategy. 

\paragraph{Learnable task weights.}
Two scalar task weights $(w_\mu,w_\sigma)$ are introduced to balance the data-fidelity losses $\mathcal{L}_\mu$ and $\mathcal{L}_\sigma$, and their associated gradients, in the total objective. To ensure positivity of the loss weights and avoid degenerate solutions, these weights are parameterised using a softmax normalisation,
\begin{equation}
(w_\mu,w_\sigma) = 2\,\mathrm{softmax}(s_\mu,s_\sigma),
\end{equation}
which enforces the constraint $w_\mu + w_\sigma = 2$ throughout training. The scalars $(s_\mu, s_\sigma)$ are optimised jointly with the network parameters.

\paragraph{GradNorm update rule.}
GradNorm aims to balance the optimisation speed of the different tasks by equalising their gradient magnitudes. Let $\theta$ denote the subset of network parameters on which the weighting strategy is applied. In this study, $\theta$ corresponds to the output layer only, i.e. the parameters of the final convolution layer producing
$\widehat{\mu}(x,y)$ and $\widehat{\sigma}(x,y)$. Restricting GradNorm to the output layer stabilises training while preserving its ability to balance learning rates between tasks. For each task $i\in\{\mu,\sigma\}$, we respectively compute the $L_2$ norm of the weighted gradients $G_i(t)$ and the average gradient norm across the tasks $\overline{G}(t)$ at training time $t$
\begin{equation}
G_i(t) = \left\lVert \nabla_{\theta}\big(w_i(t)\,\mathcal{L}_i(t)\big)\right\rVert_2,
\qquad \overline{G}(t) = \frac{1}{2}\sum_{i\in\{\mu,\sigma\}} G_i(t).
\label{eq:gradnorm_G}
\end{equation}
Let $\mathcal{L}_i(0)$ denote the initial task losses recorded at the first optimisation step, and $r_i(t)$ the relative inverse training rate, defined as
\begin{equation}
r_i(t) = \frac{\mathcal{L}_i(t) / \mathcal{L}_i(0)}{\frac{1}{2}\sum_{j\in\{\mu,\sigma\}} \mathcal{L}_j(t) / \mathcal{L}_j(0)},
\label{eq:gradnorm_relative_rate}
\end{equation}
which measures whether a task is learning faster or slower than average. In our implementation, GradNorm is applied to the data-fidelity losses $\mathcal{L}_\mu$ and $\mathcal{L}_\sigma$ only while the gradient-based losses are treated as auxiliary regularisation terms and are excluded from the computation of gradient norms and relative training rates. The target gradient norm for each task is then given by
\begin{equation}
G_i^{\ast} = \bar{G}\, r_i^{\alpha},
\end{equation}
where $\alpha$ controls the strength of rebalancing. The task weights are then updated by minimising the GradNorm penalty
\begin{equation}
L_{\mathrm{GN}}(t) = \sum_{i\in\{\mu,\sigma\}}
\left|\,G_i(t)-G_i^{\star}(t)\,\right|,
\label{eq:gradnorm_penalty}
\end{equation}
implemented using an $L_1$ loss. In practice, this update is applied every five optimisation steps to stabilise weight dynamics.

\paragraph{Hyperparameter selection.}
We evaluated several combinations of the hyperparameters $\alpha$ and $\lambda$ (see Eq.~\ref{eq:loss_function}) to assess their impact on training stability and loss balancing. The values $\alpha=0.1$ and $\lambda=1.0$ yielded the most stable convergence and consistent training behaviour across runs. This setting was therefore selected for the final U-Net model, whose results are presented in Section~\ref{sec:Results}. It provides effective balancing between the two regression objectives while avoiding manual tuning of their relative importance.

\section{Supplementary results for natural fracture \#2}
\label{app:NF2}

This appendix provides detailed fracture statistics, uncertain permeability fields, and upscaling results for the second natural shear fracture, complementing the main analysis presented in Sections~\ref{subsec:fracture_comparison} and ~\ref{sec:Results}.

\begin{figure}[ht!]
\centering
\begin{subfigure}{.49\textwidth}
    \caption{}
    \centering
    \includegraphics[width=\linewidth]{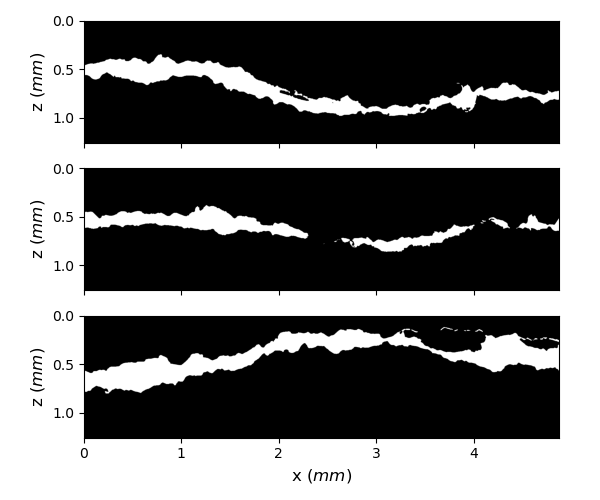}
    \label{fig:cross_sec_real_NF2}
\end{subfigure}
\hspace{-1mm}
\begin{subfigure}{.49\textwidth}
    \caption{}
    \centering
    \includegraphics[width=\linewidth]{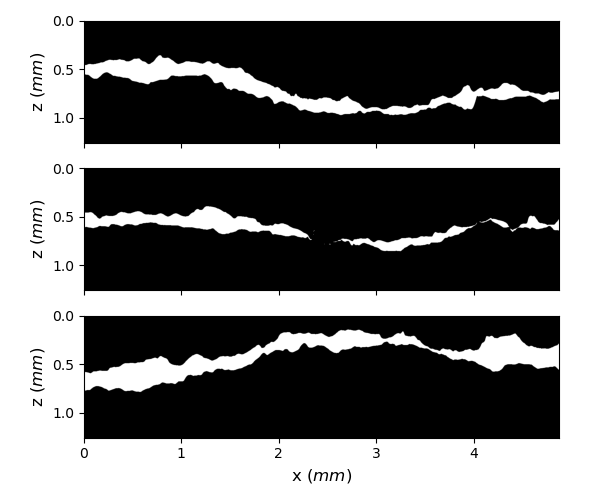}
    \label{fig:cross_sec_synthetic_NF2}
\end{subfigure}
\caption{Representative segmented cross-sections extracted along the $y$-axis for (a) the natural shear fracture \#2 and (b) its synthetic counterpart. Compared to the natural fracture \#1 (see Figure \ref{fig:cross_sec}), fracture \#2 exhibits fewer branching structures. The synthetic geometry preserves the main aperture envelope while collapsing secondary branches into a single connected channel.}
\label{fig:cross_sec_NF2}
\end{figure}

\begin{figure}[t!]
\centering
\begin{subfigure}{.42\textwidth}
    \caption{}
    \centering
    \includegraphics[width=\linewidth]{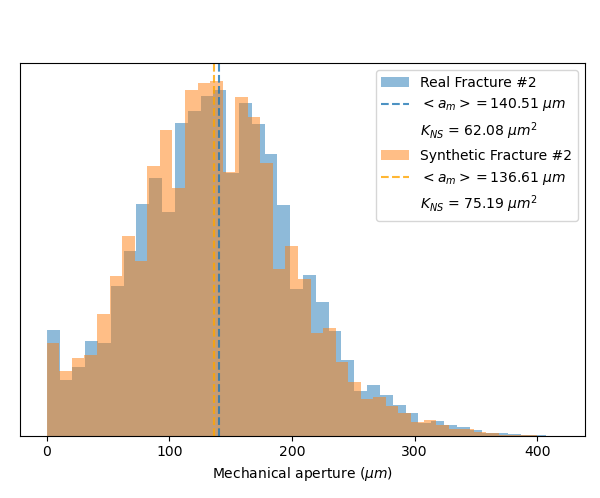}
    \label{fig:synthetic_dist_NF2}
\end{subfigure}
\hspace{-1mm}
\begin{subfigure}{.56\textwidth}
    \caption{}
    \centering
    \includegraphics[width=\linewidth]{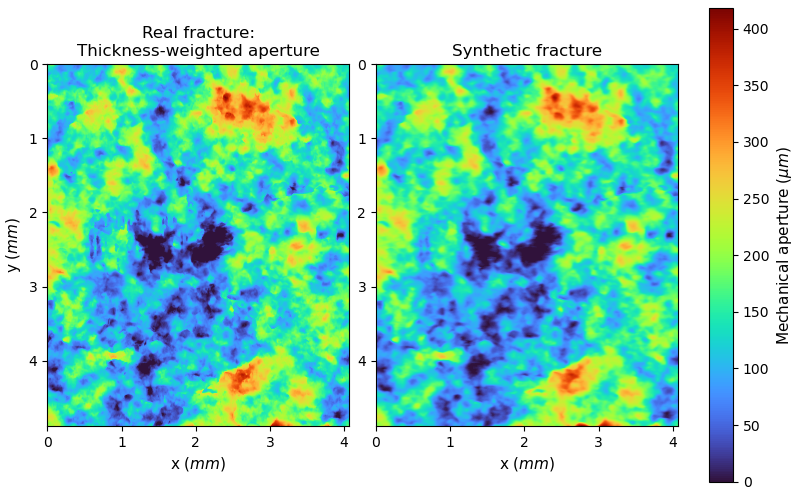}
    \label{fig:aperture_maps_NF2}
\end{subfigure}
\caption{(a) Mechanical aperture distributions for natural shear fracture \#2 and its synthetic counterpart. The mean mechanical aperture, denoted $\langle a_m \rangle$, and the Stokes-based permeability $K_{NS}$ are reported for each configuration. Both distributions exhibit similar trend, reflecting comparable aperture variability. (b) Spatial distribution of the thickness-weighted mechanical aperture for the natural and synthetic fractures \#2. The main large-scale aperture structures are preserved in the synthetic geometry, while small-scale heterogeneities and branching features are reduced, leading to a more spatially continuous aperture field.}
\label{fig:real_vs_synthetic_apertures_NF2}
\end{figure}

\begin{figure}[ht!]
    \centering
    \includegraphics[width=\linewidth]{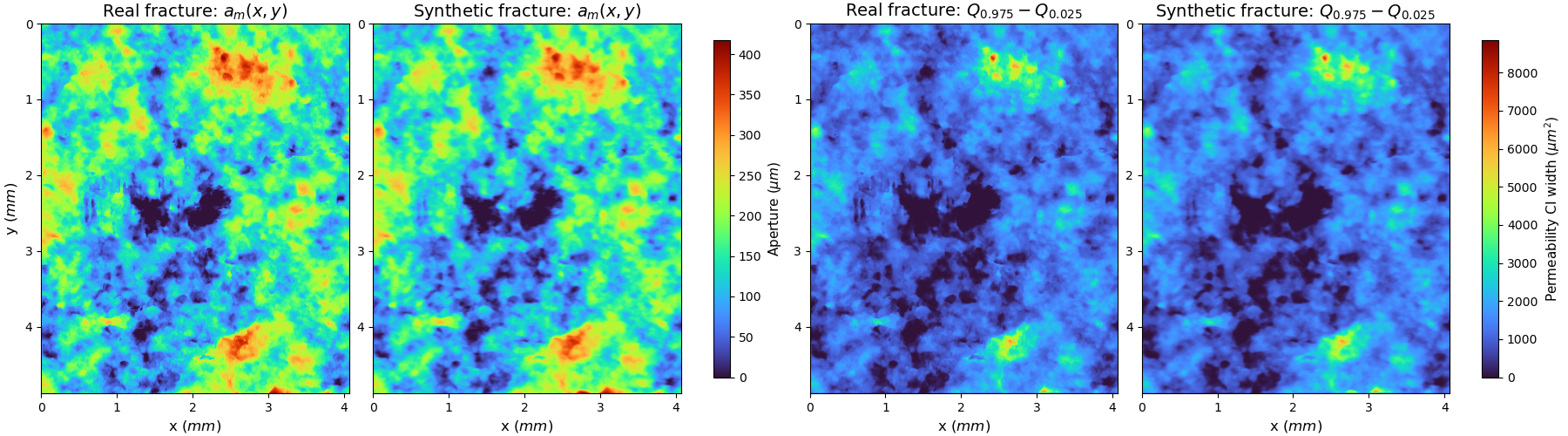}
    \caption{Thickness-weighted mechanical aperture fields $a_m(x,y)$ and corresponding local permeability uncertainty widths, quantified as the difference between the upper and lower quantiles for the natural fracture \#2 and its synthetic counterparts. }
\label{fig:Ap_vs_UQ_width_NF2}
\end{figure}

\begin{figure}[ht!]
    \centering
    \includegraphics[width=0.8\linewidth]{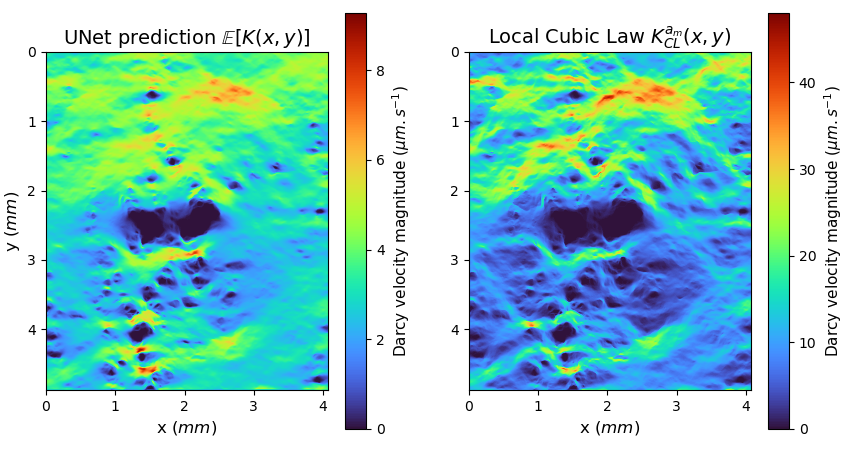}
    \caption{
    Comparison of Darcy velocity magnitude fields for the natural shear fracture \#2 obtained using two permeability models:
    (a) Velocity field computed from the probabilistic permeability predicted by the U-Net, using the expected value $\mathbb{E}[K(x,y)]$ of the local log-normal distribution.
    (b) Velocity field computed from a deterministic permeability derived from the local Cubic-Law applied to the mechanical aperture, $K^{\,a_m}_{CL}(x,y)$. }    
\label{fig:Darcy_velocities_mag_NF2}
\end{figure}

\begin{figure}[hpbt]
\centering
\raisebox{0.18\height}{
\begin{subfigure}{.39\textwidth}
    \caption{}
    \centering
    \includegraphics[width=0.92\linewidth]{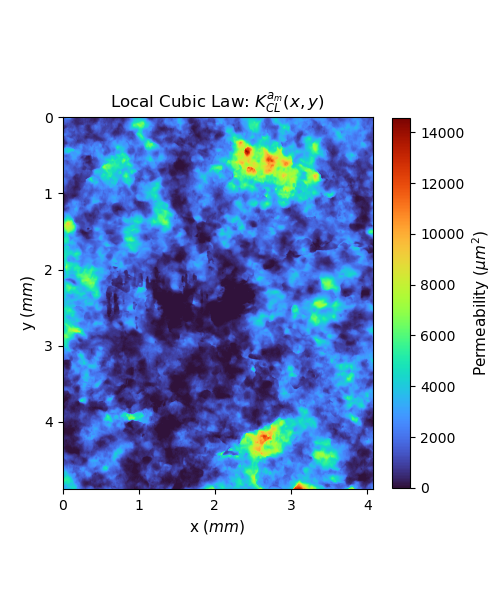}
    \label{fig:LCL_real_NF2}
\end{subfigure}
}%
\begin{subfigure}{.59\textwidth}
    \caption{}
    \centering
    \includegraphics[width=0.92\linewidth]{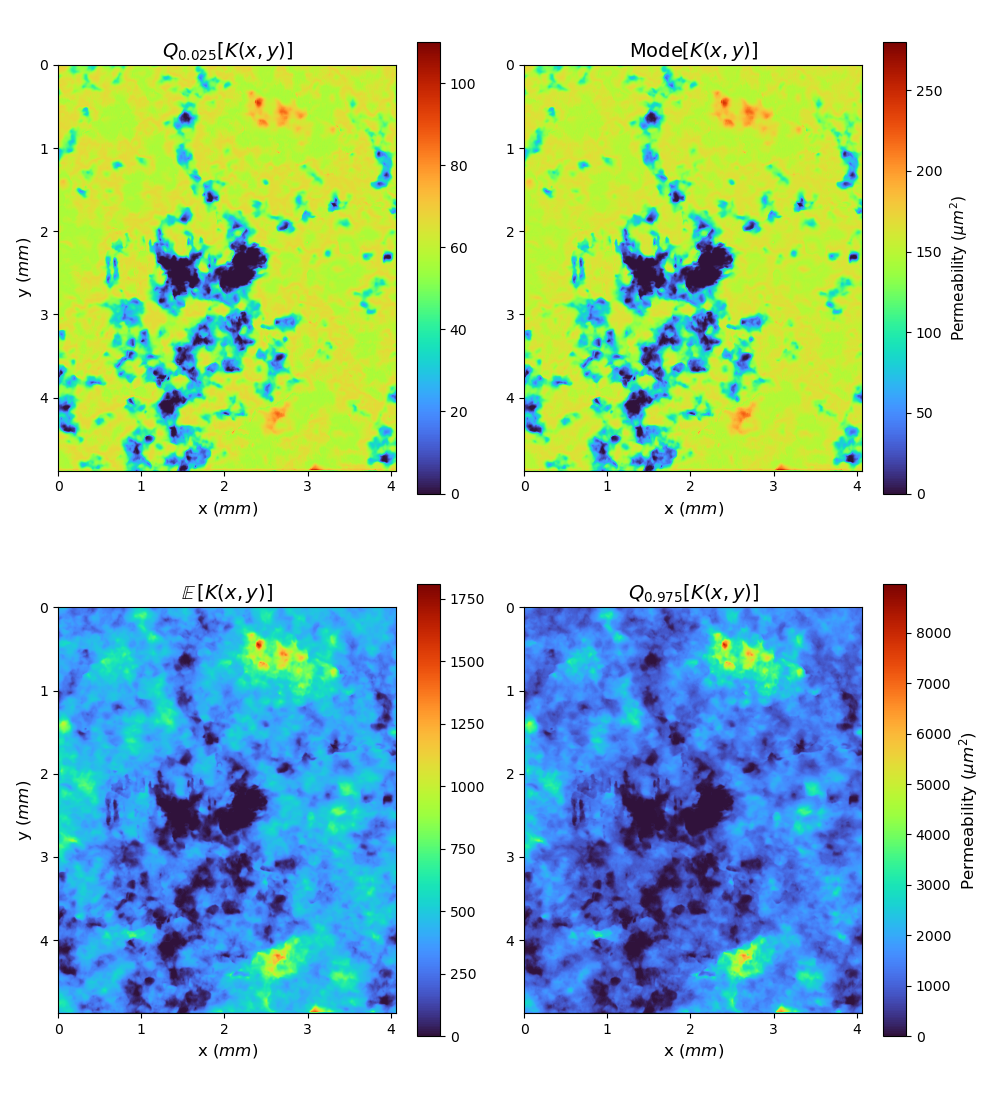}
    \label{fig:Post_K_real_NF2}
\end{subfigure}\vspace{-6mm} \rule{0.5\textwidth}{0.4pt}
\raisebox{0.18\height}{
\begin{subfigure}{.39\textwidth}
    \caption{}
    \centering
    \includegraphics[width=0.92\linewidth]{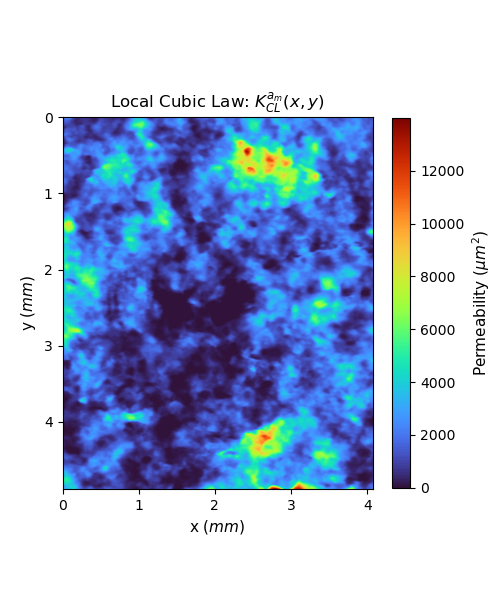}
    \label{fig:LCL_synthetic_NF2}
\end{subfigure}
}%
\begin{subfigure}{.59\textwidth}
    \caption{}
    \centering
    \includegraphics[width=0.92\linewidth]{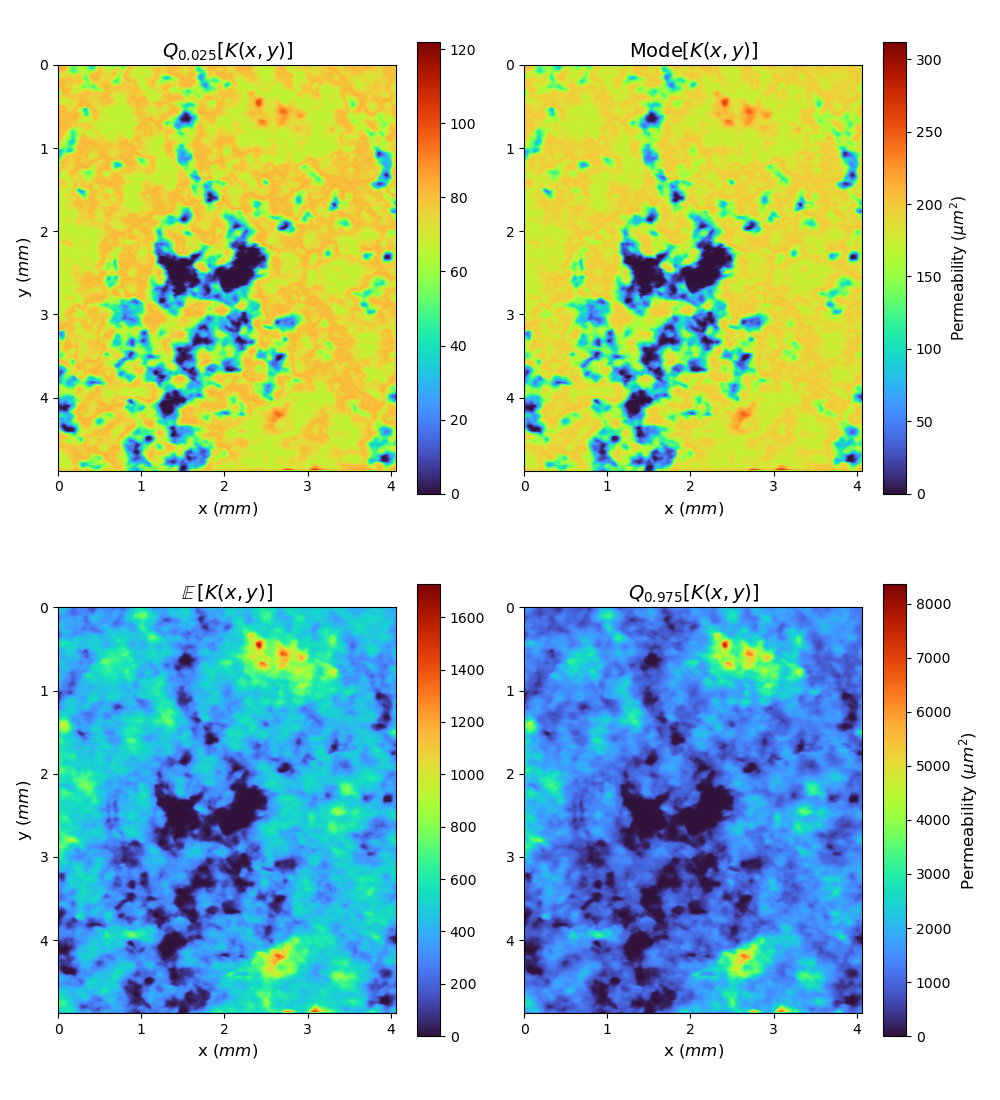}
    \label{fig:Post_K_synthetic_NF2}
\end{subfigure}
\caption{Probabilistic permeability maps for the natural shear fracture \#2 and its synthetic counterpart: (a, c) Deterministic permeability fields obtained from the local Cubic-Law, $K_{CL}^{a_m}(x,y)$. (b, d) Probabilistic permeability descriptors predicted by the U-Net, including the lower quantiles, modes, expected values, and upper quantiles.}
\label{fig:real_results_K_NF2}
\end{figure}

\clearpage

\section*{Acknowledgements}
This work is funded by the Engineering and Physical Sciences Research Council's ECO‐AI Project Grant (reference number EP/Y006143/1) and NWS-NERC GeoSAFE grant (reference NE/Y002504/1). Opinions, findings, conclusions or recommendations expressed are those of the authors and do not necessarily reflect the views of NERC or NWS.

\section*{Conflict of Interest disclosure}
The authors declare there are no conflicts of interest for this manuscript.

\section*{Open Research Section}
The source code for the probabilistic upscaling workflow developed in this study, including Residual U-Net training, full-fracture inference, and probabilistic upscaling analysis, is hosted on GitHub and archived with a DOI at Perez (2026a) \cite{Perez2026_Code}. The processed training dataset and pre-trained model required to reproduce the full workflow are archived separately at Perez (2026b) \cite{Perez2026_Dataset}. Representative fracture aperture inputs are also provided to support reproducibility and to enable direct reuse of the workflow on example data. The Bayesian correction framework that underpins the present approach is archived at Perez (2025) \cite{Perez2025_Code}.

\end{document}